\documentclass[10pt,twocolumn,prl,aps,superscriptaddress,floatfix,nobalancelastpage,
preprintnumbers,reprint,hyperref]{revtex4-2}

\usepackage[utf8]{inputenc} % usually not needed (loaded by default)
\usepackage[T1]{fontenc}

\usepackage[colorlinks=true,breaklinks=true]{hyperref}
\usepackage{comment}
\usepackage[utf8]{inputenc}
\hypersetup{allcolors=[rgb]{0.0 0.0 0.6},linkcolor=[rgb]{0.75 0.05 0.05}}
\usepackage{mathtools}
\usepackage[dvipsnames]{xcolor}
\usepackage{float}
\usepackage{letltxmacro}
\LetLtxMacro{\oldcite}{\cite}
\renewcommand{\cite}[1]{\mbox{\oldcite{#1}}}

\usepackage{orcidlink}

\long\def\exclude#1{}

\DeclareMathOperator{\MeV}{MeV}

\bibpunct{[}{]}{,}{n}{}{,}

\begin{document}

\title{Multimessenger Constraints on Radiatively Decaying Axions from GW170817}

\author{M.~Diamond~\orcidlink{0000-0003-1221-9475}} 
%\email{}
\affiliation{Arthur B. McDonald Canadian Astropartical Physics Institute,
Queens University, Kingston, Ontario, Canada K7L 3N6}
\author{D.~Fiorillo~\orcidlink{0000-0003-4927-9850}}%\email{}
\affiliation{Niels Bohr International Academy, Niels Bohr Institute,
University of Copenhagen, Blegdamsvej 17, 2100 Copenhagen, Denmark}
\author{G.~Marques-Tavares \orcidlink{0000-0002-1861-7936}}%\email{}
\affiliation{Maryland Center for Fundamental Physics, Department of Physics,
University of Maryland, College Park, MD 20742, U.S.A.}
\author{I.~Tamborra~\orcidlink{0000-0001-7449-104X}}%\email{}
\affiliation{Niels Bohr International Academy, Niels Bohr Institute,
University of Copenhagen, Blegdamsvej 17, 2100 Copenhagen, Denmark}
\affiliation{DARK, Niels Bohr Institute,
University of Copenhagen, Jagtvej 128, 2200 Copenhagen, Denmark}
\author{E.~Vitagliano~\orcidlink{0000-0001-7847-1281}}%\email{edoardo@physics.ucla.edu}
\affiliation{Racah Institute of Physics, Hebrew University of Jerusalem, Jerusalem 91904, Israel}

\date{\today}

\begin{abstract}
The metastable
hypermassive neutron star produced in the coalescence of two neutron stars can copiously produce axions  that  radiatively decay into $\mathcal{O}(100)$~MeV photons. These photons can form a fireball with characteristic temperature smaller than $1\rm\, MeV$. By relying on X-ray observations of GW170817/GRB 170817A 
with CALET CGBM, Konus-Wind, and Insight-HXMT/HE, 
 we present new bounds on the axion-photon coupling for axion masses in the range $1$--$400\,\rm MeV$. We exclude couplings down to $5\times 10^{-11}\,\rm GeV^{-1}$, complementing and surpassing existing constraints. 
Our approach can be extended  to any feebly-interacting  particle  decaying into photons.

\end{abstract}

\maketitle

\textit{Introduction.}---The first observation of a binary neutron star (NS) merger event in gravitational waves and electromagnetic radiation, GW170817, has shed new light on the  properties of NSs, the behavior of matter at nuclear densities, as well as the synthesis of the elements heavier than iron~\cite{LIGOScientific:2017ync,LIGOScientific:2017zic,Burns:2019byj}. Besides providing crucial insights on fundamental physics, NS mergers can be employed as laboratories to test physics beyond the Standard Model, such as long-range interactions and general relativity modifications (e.g.~\cite{Berti:2015itd,Ezquiaga:2017ekz,Baker:2017hug,Boran:2017rdn,Creminelli:2017sry,Langlois:2017dyl,Sakstein:2017xjx,Sennett:2019bpc,Sagunski:2017nzb,Croon:2017zcu,Huang:2018pbu,Dror:2019uea,Hook:2017psm,Zhang:2021mks}), and their remnant can produce light axions~\cite{Dietrich:2019shr,Fiorillo:2021gsw}, sterile neutrinos~\cite{Sigurdarson:2022mcm}, 
as well as dark photons~\cite{Diamond:2021ekg}. Moreover, the compact object resulting from the merger can potentially provide new and complementary information on
putative heavy particles beyond the Standard Model, which could have an impact on cosmology~\cite{Cadamuro:2011fd,Depta:2020zbh,Kelly:2020aks,Langhoff:2022bij} or play
the role of dark matter mediator~\cite{Pospelov:2007mp,Knapen:2017xzo}, and that cannot be excluded through the cooling of stars like horizontal branch stars, red giants, or white dwarfs~\cite{Raffelt:1996wa,Raffelt:2006cw}. Being hot and dense, the remnant can produce particles with mass $\gtrsim 1\,\rm MeV$, akin to  core-collapse supernovae (SNe) and other energetic transients~\cite{Jaeckel:2017tud,Chang:2018rso,Sung:2019xie,Lucente:2020whw,Croon:2020lrf,Caputo:2021rux,Caputo:2022mah,Caputo:2022rca,Fiorillo:2022cdq,Hoof:2022xbe,Diamond:2023scc,Muller:2023vjm,Caputo:2021kcv}.
\begin{figure}[t!]
    \centering
        \includegraphics[width=0.43\textwidth]{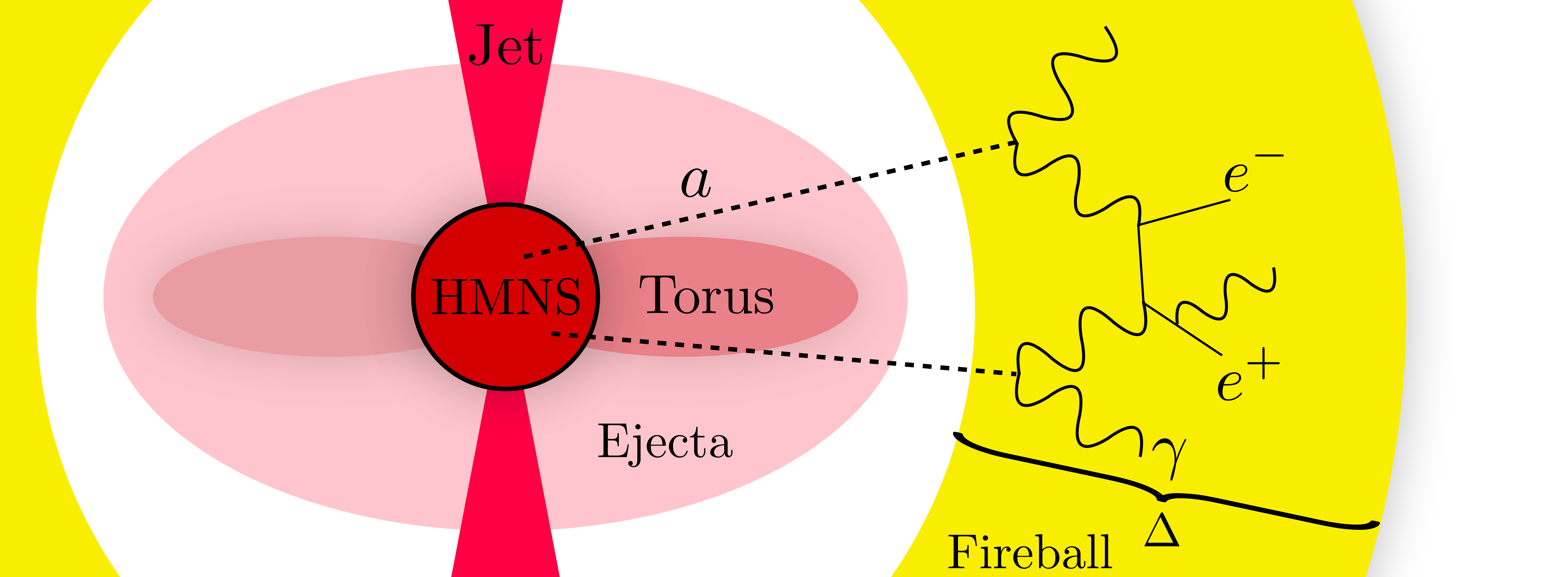}
\centering
    \includegraphics[width=0.46\textwidth]{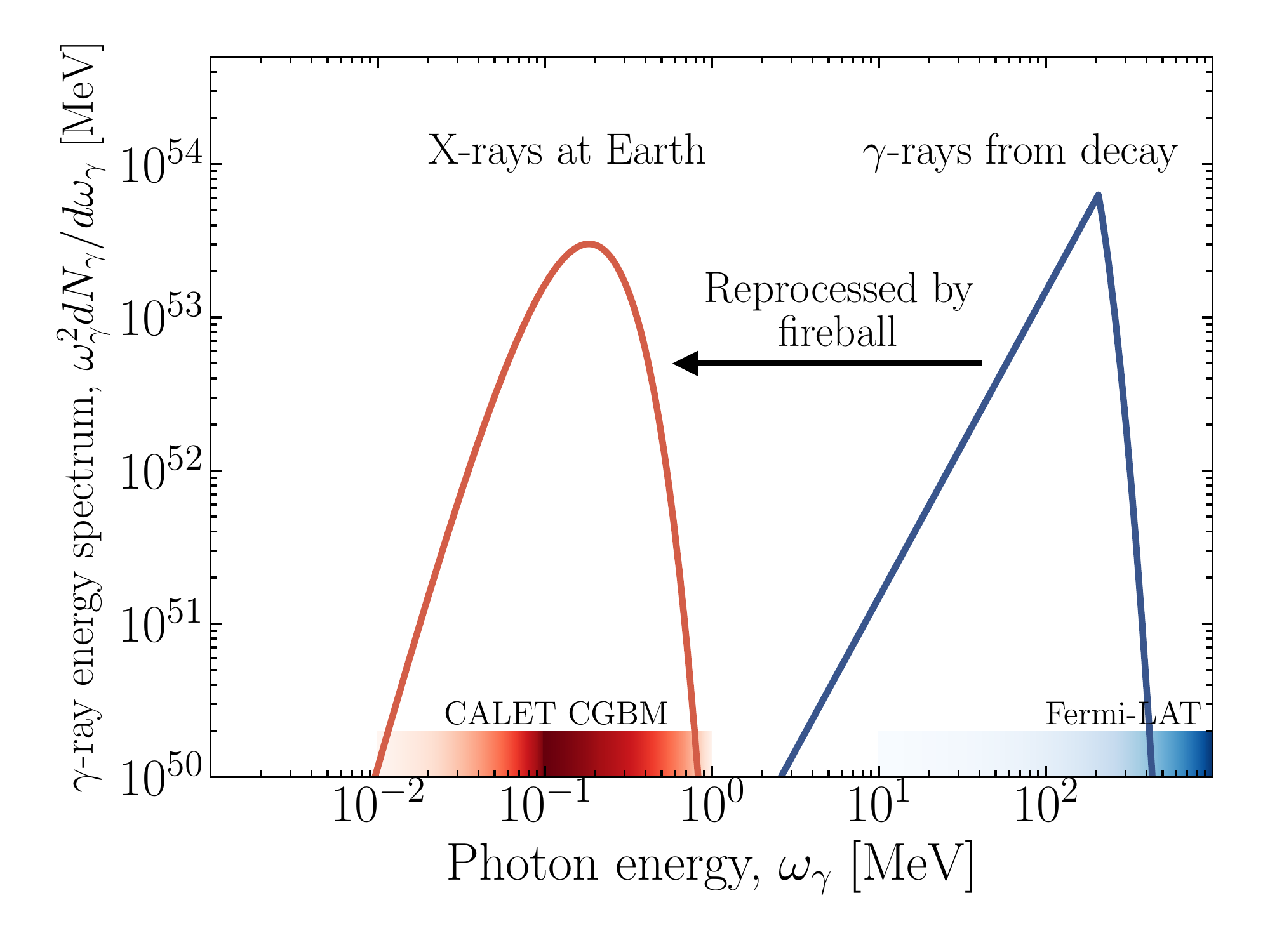}
    \caption{{\it Upper panel}: Schematic representation of the NS merger remnant and the fireball produced from axion decay.
   {\it Lower panel}: $\gamma$-ray spectrum produced by axion decay (blue line) and final spectrum reprocessed by the fireball (red line). The photon energy is reduced from $100\,\rm MeV$ to less than $1\,\rm MeV$; the color bars on the bottom show the sensitivity ranges of Fermi-LAT and CALET CGBM, which we use to set bounds. Axion mass and coupling are set to $m_a=202$~MeV and $g_{a\gamma\gamma}=2.2\times10^{-10}$~GeV$^{-1}$ for illustrative purposes.}
    \label{fig:reprocessing}
\end{figure}

The multimessenger signals of GW170817 are consistent with the formation of 
a metastable hypermassive NS (HMNS) which lived for up to $1\,\rm s$~\cite{LIGOScientific:2017vwq,Murguia_Berthier_2021} (see later discussion for implication of the  lifetime of the remnant), before collapsing into a black hole (BH). We show that one can probe the  production of heavy axion-like particles with mass up to several hundreds MeVs  with  coupling to photons $-\frac{1}{4}g_{a\gamma\gamma}a F\Tilde{F}$ (axions for short) in the HMNS remnant. 
After being produced, axions leave the HMNS and decay radiatively into high-energy ($\simeq 100\,\rm MeV$) photons, as sketched in the top panel of Fig.~\ref{fig:reprocessing}. Since we focus on heavy semi-relativistic axions, the daughter photons are dense enough that they do not propagate freely. Rather, they interact with each other rapidly producing a fireball, a plasma shell with temperature $\simeq 100 \,\rm keV$ in the HMNS remnant frame, as we have recently pointed out in the context of SNe in Ref.~\cite{Diamond:2023scc}. This gas later evolves similarly to a ``standard'' fireball propagating in vacuum. 
Differently from the fireball assumed to power gamma-ray bursts (see e.g.~Refs.~\cite{Piran:1999kx,Meszaros:2006rc}), the axion-sourced fireball features little-to-no baryon loading, is not expected to accelerate non-thermal particles, and forms almost instantaneously after the NS merger (hence the time of fireball formation can be inferred from gravitational wave observations). The fireball first expands adiabatically and then freely. The resulting photons reach Earth with a quasi-thermal spectrum with low average energies.

Crucially, the signal arising from axions with a relatively short life-time produced in a NS merger  consists of reprocessed photons that travel to Earth, and it should therefore be detected by X-ray detectors, rather than, as one may naively expect, $\gamma$-ray detectors such as Fermi-LAT~\cite{Ajello:2018mgd} (see lower panel of Fig.~\ref{fig:reprocessing}). In this {\it Letter}, we present novel bounds on axions from the non-observation of an axion-sourced fireball at GW170817/GRB 170817A by CALET CGBM~\cite{CALET}, Konus-Wind~\cite{Konus}, and Insight-HXMT/HE~\cite{Insight}.

\textit{Reference neutron star merger remnant model for particle emission.}---Observations of GW170817 suggest an asymmetric NS merger with  primary mass of $1.36$--$1.89\,M_\odot$ and  secondary mass of $1.00$--$1.36\,M_\odot$ assuming high spin (respectively, $1.36$--$1.60\,M_\odot$ and $1.16$--$1.36\,M_\odot$ for low spin scenarios)~\cite{LIGOScientific:2018hze}. In order to compute the axion production rate, we rely on the suite of binary NS merger remnant models presented in Refs.~\cite{Ardevol-Pulpillo:2018btx,George:2020veu,JankaWeb,phdthesisPulpillo}
and obtained though  three-dimensional relativistic  particle hydrodynamic simulations (see Refs.~\cite{Ardevol-Pulpillo:2018btx,George:2020veu} for more details). 

First, we  consider as our fiducial model the simulation of two non-rotating NSs with mass of  $1.45 \,M_\odot$ and $1.25\,M_\odot$, respectively, and nuclear equation of state (EoS) SFHo. In the HMNS core, where axion production occurs, the typical temperature is a few tens of~MeV and the baryon density is around $10^{14}\,\rm g/cm^3$. The benchmark simulation we use tracks the NS merger remnant evolution up to $10\,\rm ms$, and we assume~(in agreement with results from Refs.~\cite{Sekiguchi:2011zd,Camelio:2020mdi} that the remnant has reached a steady state and does not appreciably changes its thermodynamical properties up to $1\,\rm s$, namely up to the time considered for BH formation; we later investigate the impact of this assumption on the axion bounds. Then, we also  compute the uncertainty introduced by the EoS and the mass of the two NSs  by considering another EoS (DD2), and different NS masses (symmetric merger model, with  two NSs of $1.35\,M_\odot$ mass)~\cite{Ardevol-Pulpillo:2018btx,George:2020veu,phdthesisPulpillo}.

\textit{Axion and photon spectra.}---Axions are produced in the HMNS mainly via two different processes. One is the Primakoff effect, i.e.~photons that convert into axions in the field generated by charged particles ($\gamma+Ze\rightarrow a+Ze$), while the other is photon coalescence
($\gamma+\gamma\rightarrow a$)~\cite{Lucente:2020whw,Caputo:2021rux,Caputo:2022mah}. We obtain the axion spectrum integrating over the volume of the HMNS  and time,
\begin{align}\label{eq:axionspectrum}
        \frac{dN_a}{d \omega_a}=\ &\frac{1}{2\pi^2}\int dV dt \frac{1}{e^{ \omega_a/T}-1}\nonumber \\
       & \times\left(\Gamma_{\rm P} \omega_a\sqrt{ \omega_a^2-\omega_{\rm P}^2}+\Gamma_{\rm c} \omega_a\sqrt{ \omega_a^2-m_a^2}\right)\ ,
\end{align}
where $\Gamma_{\rm P}$ and $\Gamma_{\rm c}$ are the Primakoff and coalescence production rates, and $\omega_{\rm P}$ is the plasma frequency modifying the photon dispersion relation inside the HMNS, $\omega^2=k^2+\omega_{\rm P}^2$.  
We account for gravitational redshift correction of  the energy. We refer the interested reader to the  Supplemental Material for additional details~\cite{supplementalmaterial}.
Axions subsequently decay into photons away from the HMNS, at a distance of $\mathcal{O}(10^{3}$--$10^6$)~km. 
The photon spectrum right after the axion decay (and before photons interact with each other) is easily found assuming a box spectrum for the daughter photons~\cite{Oberauer:1993yr,Jaffe:1995sw,Caputo:2021rux},
\begin{align}
        \frac{dN^{i}_{\gamma}}{d\omega_{\gamma}}=2\int_{\omega_\gamma}^\infty\frac{dN_a}{d \omega_a}\frac{d \omega_a}{ \omega_a}\,
\end{align} 
where $i$ stands for ``initial''.

\begin{figure}
    \centering
    \includegraphics[width=0.49\textwidth]{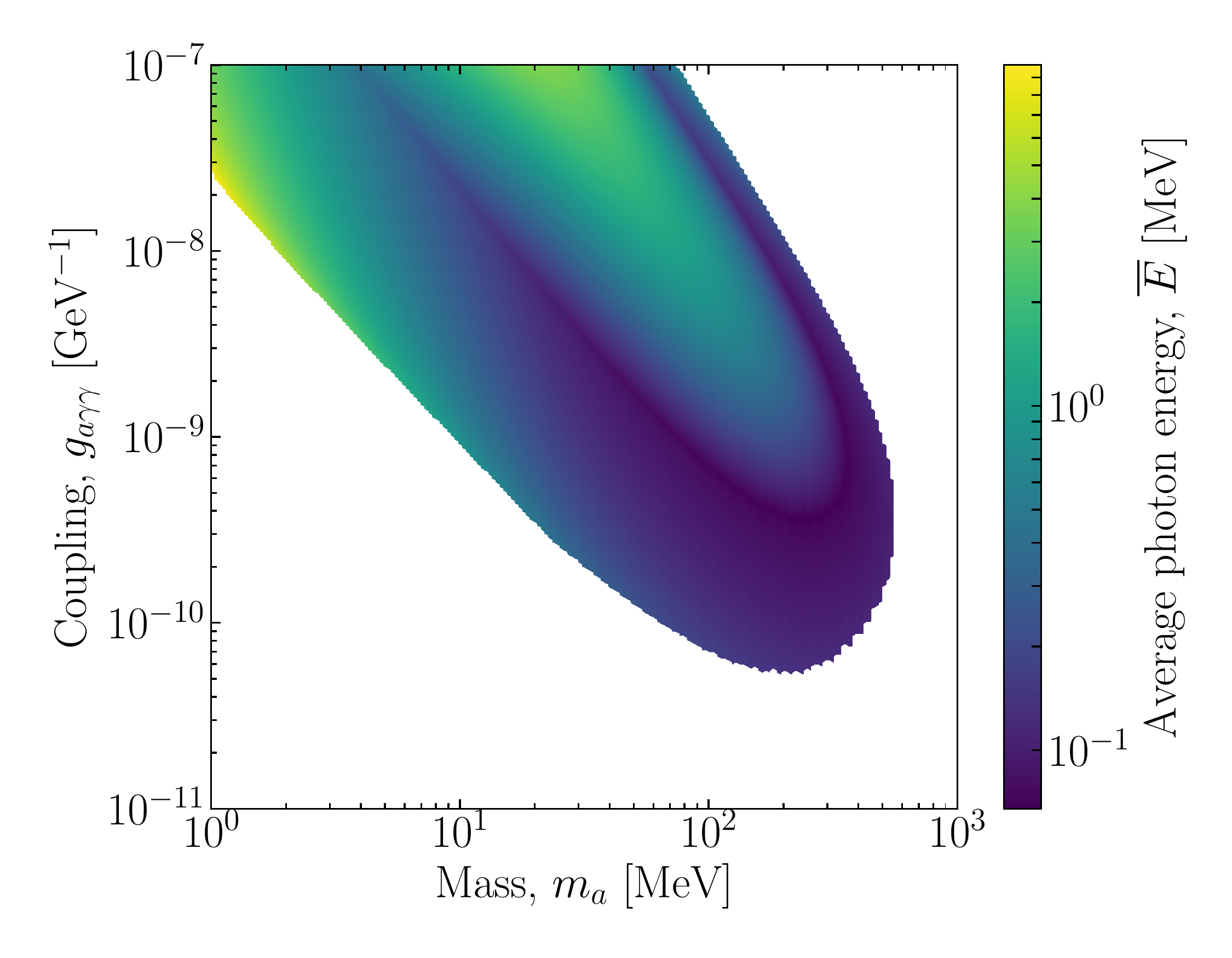}
    \caption{Isocontours of the average photon energy at the end of the fireball evolution, in the plane spanned by  the axion mass and coupling. In the white region, the fireball does not form. As the coupling increases, the average energy first lowers, due to more efficient bremsstrahlung, and then increases, due to the larger axion production and the smaller fireball radius. 
    }
    \label{fig:average_energy}
\end{figure}

\textit{Fireball production.}---If the injected photons are dense enough, they form a shell of thermalized photon fluid which we dub a fireball, diluting the photon average energy to the sub-MeV range. The physics behind this process is described in Ref.~\cite{Diamond:2023scc}, which we refer to for  technical details. We model  injection as a uniform shell of photons produced by the decay of axions and denote the shell radius with $r$, and the shell thickness $\Delta$. For each axion mass and coupling, the fireball properties are self-consistently determined following Ref.~\cite{Diamond:2023scc}, accounting for only those axions decaying outside a minimum radius of $1000$~km, below which photons free escape would be impeded~\cite{Diamond:2021ekg}. Fireball formation requires both pair production, to produce seed electron-positron pairs, and the subsequent bremsstrahlung reaction of $e^\pm$, to increase the number of particles via $e\to e\gamma$, to be fast enough. With this criterion, we identify the region of parameter space in which the fireball can form.

Photons initially thermalize with a large chemical potential $\mu_{\gamma,i}<0$ and an initial temperature of the order of the axion mass $T_{\gamma,i}$; the electron and positron populations both have the same chemical potential and temperature. As bremsstrahlung proceeds, the average energy per particle is diluted, reducing both $|\mu_\gamma|$ and $T_\gamma$; as $T_\gamma$ becomes smaller than the electron mass, the $e^\pm$ population in the plasma is depleted by pair annihilation. If it becomes sufficiently rarefied, bremsstrahlung stops, with the plasma temperature determined by the freeze-out condition
\begin{equation}\label{eq:bremsstrahlung_equilibrium}
    \gamma n_e(T_\gamma,\mu_\gamma) v_\mathrm{th} \sigma_{ee\to ee\gamma} \Delta=1\ ,
\end{equation}
where $n_e(T_\gamma,\mu_\gamma)$ is the electron number density, $v_\mathrm{th}$ is the thermal velocity, and $\sigma_{ee\to ee\gamma}$ is the bremsstrahlung cross section, and $\gamma$ the  Lorentz factor.
If the plasma is dense enough, bremsstrahlung may completely equilibrate the plasma, in which case the final state is rather determined by the condition $\mu_\gamma=0$. In addition, conservation of the total energy $\mathcal{E}$ and radial momentum $\mathcal{P}$ of the plasma must be enforced. These three conditions together determine the final temperature $T_\gamma$, chemical potential $\mu_\gamma$, and Lorentz factor $\gamma$ of the fireball. Finally,  the spectrum observed at Earth is~\cite{Diamond:2023scc}
\begin{equation}\label{eq:spectrum}
    \frac{dN}{dE}\propto -E \log\left[1-e^{-\eta-\frac{E}{2\tau}}\right],
\end{equation}
with $\eta=-\mu_\gamma/T_\gamma$, $\tau=\gamma T_\gamma$, and the spectrum being normalized according to the total energy injected.

Figure~\ref{fig:average_energy} shows the average energy  $\overline{E}=4\tau \mathrm{Li}_4(\eta)/\mathrm{Li}_3(\eta)$---where $\mathrm{Li}_s(z)$ is the polylogarithm of order $s$---of the photons observed at Earth in the region of fireball formation. For a given mass $m_a$, increasing the coupling first lowers the average energy, since bremsstrahlung becomes more effective in increasing the particle number; however, at some point bremsstrahlung manages to enforce chemical equilibrium $\mu=0$, after which increasing the coupling only increases the total energy density injected by the axions and therefore also the average energy.  For  large couplings, most axions decay within the inner optically thick region without forming a fireball. Overall, the typical photon energy in the fireball is below MeV.

\textit{Axion constraints.}---We  now compare our predicted axion spectra   with the data collected by CALET CGBM~\cite{CALET}, Konus-Wind~\cite{Konus}, and Insight-HXMT/HE~\cite{Insight} from GW170817/GRB 170817A (see Table~\ref{tab:limits})~\footnote{Additional bounds may be obtained from e.g. the Fermi GBM data~\cite{Goldstein:2017mmi}, which however provide upper bounds that are comparable with the ones we use here.}. These three experiments were online on August 17th, 2017 (at 12:41:04 UTC). Since it is estimated that GW170817/GRB 170817A occurred at a distance of $D_L=41^{+16}_{-12}\,\rm Mpc$ assuming high spin or $D_L=39^{+7}_{-14}\,\rm Mpc$ for low spin~\cite{LIGOScientific:2018hze}, the upper limits on the X-ray emissivity correspond to an integrated luminosity of $3\times 10^{46}\,\rm erg$. We obtain novel stringent bounds on the axion coupling to photons by requiring that the photon fluence at Earth, integrated with the energy spectrum of Eq.~\ref{eq:spectrum} over the sensitivity interval of each experiment, is smaller than the upper limit found by X-ray telescopes, excluding part of the parameter space where an axion-sourced fireball can form.

\begin{figure*}
    \centering
    \includegraphics[width=\textwidth]{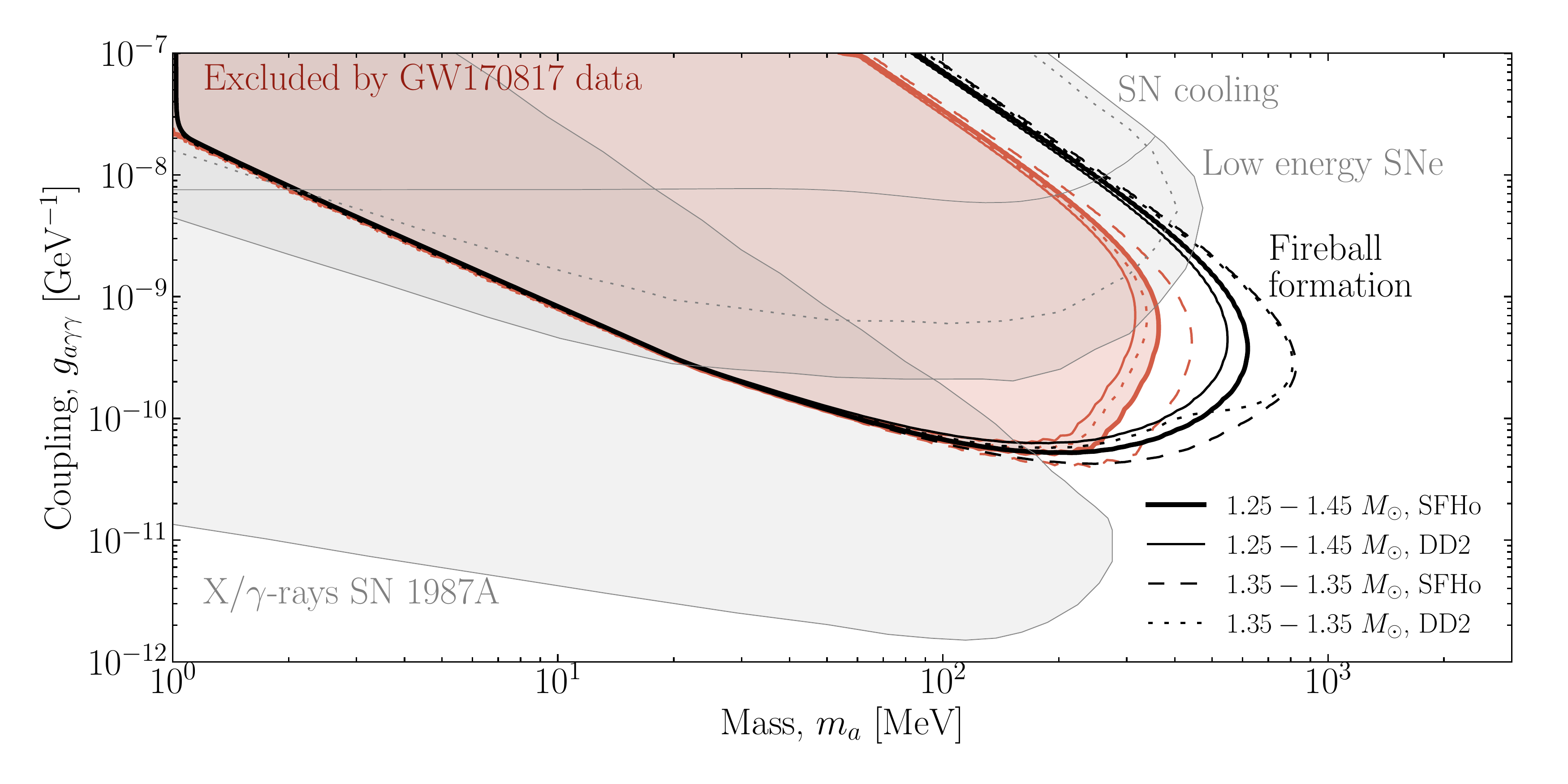}
    \caption{Constraints on the axion mass and coupling, obtained by investigating under which conditions fireball formation occurs (black lines; below $1$~MeV no fireball formation can occur since pair production cannot happen).
    The  bounds due to the non-observation of an axion-sourced fireball from  GW170817/GRB 170817A are shown in red. 
    For comparison,  the SN 1987A cooling bounds~\cite{Caputo:2021rux,Caputo:2022mah}, bounds from
low-energy SNe~\cite{Caputo:2022mah} (dotted and solid lines for conservative and fiducial bounds respectively), $\gamma$-ray~\cite{Hoof:2022xbe} and X-ray~\cite{Diamond:2023scc} bounds due to axion decays from SN 1987A are also shown. The thick and thin solid, dashed and dotted contours have been obtained for our two different EoS, as well as for symmetric and asymmetric NS merger remnant models. The non-observation of an axion-sourced fireball from  GW170817/GRB 170817A excludes a new region of the parameter space, complementary to the one excluded from core-collapse SNe. 
}
    \label{fig:bounds}
\end{figure*}

Even with just a single NS merger event, we can exclude novel parts of the axion parameter space (red region in Fig.~\ref{fig:bounds}). While the decay of axions was proposed as a mechanism to produce the fireball powering gamma-ray bursts~\cite{Berezhiani:1999qh}, this would require luminosities above $10^{52}\,\rm erg$, in conflict with low energy SNe and GW 170817 observations.

\begin{table}
 \caption{X-ray upper limits from the telescopes  online during GW170817/GRB 170817A. All the quoted upper limits are at $90\%$ C.L. and are taken from Ref.~\cite{LIGOScientific:2017ync}.}
 \vskip4pt
    \centering
    \begin{tabular*}{\columnwidth}{@{\extracolsep{\fill}}lll}
    \hline\hline
     Telescope &  Flux Upper Limit     & Energy band        \\
      & ($\rm erg \,cm^{-2}\, s^{-1} $)    &       \\
     \hline
      CALET CGBM & $1.3 \times 10^{-7}$ &  $10$--$1000$ keV \\
      Konus-Wind & $3.0 \times 10^{-7}$ ($\rm erg\, cm^{-2}$) &$10\,\rm keV$--$10\,\rm MeV$  \\
      Insight-HXMT/HE & $3.7 \times 10^{-7}$ &  $0.2$--$5$ MeV \\
     \hline
    \end{tabular*}\label{tab:limits}
\end{table}

\textit{One-zone model.}---The dependence of the axion bounds on the NS merger remnant model  raises the  question: what parameters of the HMNS mostly impact our bounds? To answer this question, we work out a  one-zone model showing the bound dependence on the NS merger remnant properties. We model the HMNS as a sphere with uniform temperature $T$ and radius $R$, lasting for a time $\delta t$. In the new region excluded in this work, the dominant emission process is photon-photon coalescence, so we only consider this process. The total energy injected in axions is
\begin{equation}
    \mathcal{E}=\frac{g_{a\gamma\gamma}^2 T^3 m_a^4R^3\delta t}{96 \pi^2}e^{-m_a/T}\sqrt{\frac{\pi m_a^3}{2 T^3}}\ ,
\end{equation}
while the total number of axions injected is
\begin{equation}
    \mathcal{N}=\frac{g_{a\gamma\gamma}^2T^2 m_a^4 R^3 \delta t}{96 \pi^2}e^{-m_a/T}\sqrt{\frac{\pi m_a}{2T}}\ .
\end{equation}

The excluded region is determined by two conditions; first, the fireball must form, so that photons are reprocessed in the  region below the MeV range. In the large mass region, it is sufficient that pair annihilation is fast enough. Assuming that axions decay at a typical radius equal to their rest-frame decay length, and parameterizing the pair annihilation cross section as $\sigma_{\gamma\gamma\to e^+ e^-}=8\pi\alpha^2/m_a^2 \log(m_a/m_e)$ evaluated at the typical center-of-mass energy of the photons $m_a/2$, we find:
\begin{equation}\label{eq:cond1}
    \frac{g_{a\gamma\gamma}^6m_a^8\alpha^2T^2R^3\delta t}{98304\pi^4}e^{-m_a/T}\sqrt{\frac{\pi m_a}{2T}}\log\left(\frac{m_a}{m_e}\right)>1\ .
\end{equation}
This qualitative condition determines the floor of our new bounds. The second requirement is that the total injected energy  is larger than the threshold that would have been visible at the X-ray telescopes, $\overline{\mathcal{E}}=4\pi D_L^2 \mathcal{F} \delta t$, where we estimate $\delta t=1\;\mathrm{s}$, $D_L$ is the luminosity distance, and $\mathcal{F}$ is the upper bound on the observed flux: 
\begin{equation}\label{eq:cond2}
    \frac{g_{a\gamma\gamma}^2 T^3 m_a^4 R^3 \delta t}{96\pi^2}e^{-m_a/T}\sqrt{\frac{\pi m_a^3}{2T^3}}=\overline{\mathcal{E}}\ .
\end{equation}
This condition determines the largest masses at which our new bound closes to the right in Fig.~\ref{fig:bounds}.

From these equations, we see that the main remnant parameters affecting  our new bounds are the average temperature of the HMNS ($T$),  the average space volume, and time duration of the event ($R^3 \delta t$). Notice that the bottom tail of the bounds in Fig.~\ref{fig:bounds} is determined by  Eq.~\eqref{eq:cond1} and depends very mildly on these parameters, given the strong $g_{a\gamma\gamma}^6$ dependence. The ballpark of our bounds for our suite of NS merger remnant models can be inferred by the typical values $T\simeq 18$~MeV, $R=16$~km, and $\delta t\simeq 1$~s. 

\textit{Which among these parameters are more uncertain?}---The largest uncertainty is associated to $\delta t$, the duration over which the NS merger remnant thermodynamic properties can be considered constant before BH formation.
For simplicity, we assume $\delta t \simeq 1$~s, although our benchmark NS merger remnant simulations run up to $10$~ms. On the other hand, existing work shows that the time it takes for a HMNS to collapse into a BH can be anywhere between $20$~ms and more than $1$~s~\cite{Rosswog:2001fh,Shibata:2005ss,Shibata:2006nm,Sekiguchi:2011zd,Hotokezaka:2013iia,Bernuzzi:2015opx,Radice:2018xqa,Fujibayashi:2020dvr,Kiuchi:2022nin}, depending on the EoS, NS masses, and angular momentum of the compact HMNS. As  for  GW170817/GRB 170817A, Ref.~\cite{Gill:2019bvq} presents at least two arguments in support of  $\delta t \simeq 1$~s, based on the time needed to eject enough material to power the observed optical/UV emission and on the delay time of $1.74$~s between the gravitational waves and the electromagnetic signal. Other studies on the subject  reach similar conclusions~\cite{Granot:2017tbr,Shibata:2017xdx,Metzger:2018qfl,Murguia-Berthier:2020tfs}, and also the end-to-end simulations presented in Ref.~\cite{Just:2023wtj}  support the delayed BH formation of GW170817. Yet, the delay of the electromagnetic signal is not sufficient to conclusively claim that the HMNS lasted for $1$~s; in fact the prompt $\gamma$-ray emission may have been produced by the shock breakout driven by the circumstellar material~\cite{Gottlieb:2017pju}. Even in this case, a delay between the merger and jet breakout should have been of the order of about $1$~s, so the collapse should still have happened after about $700$~ms. For the sake of simplicity, in the following,  we assume the temperature to be constant between $10$~ms and the time of BH formation, as found in numerical simulations, see, e.g., Refs.~\cite{Sekiguchi:2011zd,Camelio:2020mdi}. Notice that even if the collapse happened earlier than $1$~s our bounds would not suffer significantly: our one-zone model shows that the floor of the bound would be weaker by a factor $(\delta t/1~\mathrm{s})^{1/6}$. The right boundary of the excluded region would weaken at most by a factor $(\delta t/1~\mathrm{s})^{1/2}$, following the one-zone model. However,  it is the highest temperatures that determine the right boundary, and such temperatures are reached in the first $10$~ms; thus, the change is  mild.

The thermodynamic properties of our benchmark NS merger remnant simulations are  conservative. Existing models, e.g.~the ones of Ref.~\cite{Camelio:2020mdi,Radice:2021jtw}, reach peak temperatures several times larger than the ones assumed here, e.g.~up to $\mathcal{O}(100)$~MeV. Therefore, axion emission could be even substantially larger than our estimate and extend to larger axion masses. On the other hand, the trapping of the fireball by the ejecta expelled after the merger can impact the chances of successfully detecting the fireball, as discussed in the Supplemental Material~\cite{supplementalmaterial}.
Since these two effects go in opposite directions, we  conclude that our results fall in the right ballpark.

\textit{Discussion and outlook.}---Multimessenger observations of NS merger remnants provide us with the unique chance to constrain the physics of feebly interacting particles decaying radiatively. We compute the electromagnetic emission due to axions produced in the HMNS resulting from the NS coalescence.
The daughter photons produced by the axions decaying after leaving the HMNS  form a shell whose temperature becomes smaller and smaller, until the gas first expands adiabatically, converting the temperature into bulk momentum, and finally expands freely. The low-energy photons ($\sim 100\,\rm keV$) produced through these mechanism should have been observed by the X-ray telescopes online at the time of the GW170817/GRB 170817A detection.
Since CALET CGBM, Konus-Wind, and Insight-HXMT/HE reported null results, we rely on their flux upper limits   to constrain the axion parameter space. Intriguingly, we place bounds for a new region of the parameter space and complement existing core-collapse SN bounds.

Our analysis  can be applied to other particles decaying into photons, such as heavy neutral leptons~\cite{Dolgov:2000jw,Fuller:2008erj,Magill:2018jla,Brdar:2023tmi}. More  precise bounds could be derived in the future once long-term sophisticated  NS merger simulations will become available.
Moreover, dedicated differential energy analyses of X-ray telescopes would improve the bounds, since the method that we have adopted  to compute the photon differential spectrum can be used to compare the predicted and  observed emissivity per energy interval. 
Finally, if upcoming observations of NS mergers should feature a very hot HMNS, it will be possible to probe axions with masses up to the GeV scale. Therefore,  future multi-messenger observations may provide us with the tantalizing opportunity of observing an axion-sourced fireball, with a quasi-thermal spectrum. Conversely, its non-observation would give further, stringent constraints on heavy axions coupling to photons.

\textbf{Note added.}---While this project was in its final stages of completion, we became aware of Ref.~\cite{Dev:2023hax} which proposes  constraints on long-lived axions from GW170817 by relying on $\gamma$-ray observations. In contrast, our paper focuses on shorter axion lifetimes (that Ref.~\cite{Dev:2023hax} does not constrain). Moreover, compared to Ref.~\cite{Dev:2023hax}, we assume a different NS merger remnant benchmark model which reaches lower temperatures, leading to more conservative axion constraints. More importantly, we account for the fireball formation; the latter allowed us to obtain novel bounds in unconstrained regions of the parameter space,  and invalidates part of the future reach projections of Ref.~\cite{Dev:2023hax}.

\textit{Acknowledgements.---}We thank Hans-Thomas Janka and Georg Raffelt for comments on a draft of this paper. MD thanks the National Sciences and Engineering Research Council of Canada (NSERC) for their support. DFGF is supported by the Villum Fonden under project no.~29388 and  the European Union’s Horizon 2020 research and
innovation program under the Marie Sklodowska-Curie
grant agreement No.~847523 ``INTERACTIONS.'' GMT acknowledges support by the National Science Foundation under Grant Number PHY-2210361 and by the US-Israeli BSF Grant 2018236.  IT thanks  the Villum Foundation (Project No.~37358), the Carlsberg Foundation (CF18-0183), and the Deutsche Forschungsgemeinschaft through Sonderforschungsbereich SFB 1258 ``Neutrinos and Dark Matter in Astro- and Particle Physics'' (NDM).  EV thanks the Niels Bohr Institute for hospitality, and acknowledges support by the European
Research Council (ERC) under the European Union’s Horizon Europe research and innovation programme (grant agreement No. 101040019) and  the Rosenfeld Foundation.

\bibliographystyle{bibi}
\bibliography{References}

\onecolumngrid
\appendix

\clearpage

\setcounter{equation}{0}
\setcounter{figure}{0}
\setcounter{table}{0}
\setcounter{page}{1}
\makeatletter
\renewcommand{\theequation}{S\arabic{equation}}
\renewcommand{\thefigure}{S\arabic{figure}}
\renewcommand{\thepage}{S\arabic{page}}

\begin{center}
\textbf{\large Supplemental Material for the Letter\\[0.5ex]
Multimessenger Constraints on Radiatively Decaying Axions from GW170817}
\end{center}

In this Supplemental Material, we provide further details about axion production in the HMNS, and discuss the impact of ejecta on the evolution of the fireball.

\bigskip

\twocolumngrid
\section{A.~Axion production}
In this appendix, we recollect results from the literature and give explicit expressions for the axion production rates appearing in the main text. We  mainly follow Refs.~\cite{DiLella:2000dn,Caputo:2021rux,Caputo:2022mah}.

For masses below tens of MeV, the dominant production mechanism is Primakoff conversion of photons scattering on charged particles, $\gamma+Ze\rightarrow a+Ze$. Assuming that the charged particle does not move (zero recoil), the Primakoff production rate is
\begin{align}
    \Gamma_{\rm P}=2\hat{n}\sigma_{\rm P}\ ,
\end{align}
where $\hat{n}$ is the charge number density and $\sigma_{\rm P}$ is the Primakoff cross section,
\begin{align}
\sigma_{\rm P}=\frac{Z^2 \alpha g_{a\gamma\gamma}^2}{2}f_{\rm P}\ .
\end{align}

The function $f_{\rm P}$ is
\begin{eqnarray}
    f_{\rm P}&=&\frac{\bigl[(k+p)^2+  k_{\rm S}^2\bigr]\bigl[(k-p)^2+  k_{\rm S}^2\bigr]}{16 k_{\rm S} ^2kp}
    \log\frac{(k+p)^2+ k_{\rm S} ^2}{(k-p)^2+  k_{\rm S}^2}
    \nonumber\\[1ex]
    &&{}-\frac{(k^2-p^2)^2}{16   k_{\rm S}^2kp}\log\frac{(k+p)^2}{(k-p)^2}-\frac{1}{4}\ ,
\end{eqnarray}
where $k$ and $p$ are the photon and axion momenta respectively, and $k_{\rm S}$ is the Debye screening scale. We approximate the HMNS plasma as being mainly composed by electrons and protons ($Z=1$). Since electrons are relativistic and degenerate, their contribution to Primakoff production and Debye screening is cumbersome to evaluate. We neglect it and assume the Debye screening scale to be
\begin{equation}\label{eq:screeningscale}
  k_{\rm S}^2=\frac{4\pi\alpha \hat n}{T}
\quad\hbox{where}\quad
  \hat{n}=Y_e n_B\ ,
\end{equation}
with $Y_e$ being the number of electrons per baryon, and $n_{\rm B}$ the baryon number density.

In a medium, photons acquire a nontrivial dispersion relation that can be roughly approximated as $\omega^2=k^2+\omega_{\rm P}^2$ (for the transverse excitations), where $\omega_{\rm P}^2=(4\alpha/3\pi)(\mu_e^2+\pi^2T^2/3)$, and $\mu_e$ is the electron chemical potential. The HMNS has typical parameters $\omega_{\rm P}\simeq10~\MeV$, to be compared with $T\simeq20\,\MeV$ and a typical photon energy $\omega\simeq 3T\simeq60\,\MeV$. Notice that the temperature peak can be much larger than this value. In the following and in the main text, we include the modifications to the photon dispersion relations. However, we stress that the low-energy tail of the spectrum (which is largely subdominant in the fireball production) should be taken with a grain of salt, as it would be also modified by a wave-function renormalization factor (see e.g.~Ref.~\cite{Raffelt:1996wa}) that we do not include in our results. Moreover, we neglect the contribution from longitudinal plasmons.

The other production channel considered in our paper is the two-photons coalescence. The production rate reads
\begin{align}
    \Gamma_{\rm c}=\frac{g_{a\gamma\gamma}^2m_a^4 f_{\rm B}}{64\pi\omega}\ ,
\end{align}
where
\begin{eqnarray}\label{eq:fBfull}
    f_{\rm B}&=&\frac{2T}{p}
     \log\left[\frac{e^{\frac{\omega+p}{4T}}-e^{-\frac{\omega+p}{4T}}}{e^{\frac{\omega-p}{4T}}-e^{-\frac{\omega-p}{4T}}}\right]\ .
\end{eqnarray}
For all production processes we account for gravitational redshifting by writing the local comoving energy as $\omega'_a= \omega_a/(1+\phi)$ where $\phi$ is the (negative) gravitational potential in dimensionless units; for each snapshot of our benchmark NS merger remnant models, we determine it by solving the Poisson's equation for the gravitational potential:
\begin{equation}
\phi(r,z)=-\int \frac{G_N \rho(r',z')}{\sqrt{(z-z')^2+r^2+r^{'2}-2r r' \cos \phi'}} r' dr' dz' d\phi'\ ,
\end{equation}
where $G_N$ is Newton's gravitational constant.
Notice that in determining $dN_a/d \omega_a$ no Jacobian factor needs to be accounted for, since the production rate $dN_a/d \omega_a dt$ is a relativistic invariant and we assume the times of the profile snapshot are measured in the Earth frame and not in the comoving frame. Finally, accounting for gravitational trapping requires to only consider particles above the escape velocity with $ \omega_a>m_a$, namely $\omega'_a>m_a/(1+\phi)$. For comparison, Ref.~\cite{Dev:2023hax} only accounts for the last requirement, but does not consider the difference between energy detected at Earth and energy at production.

\begin{figure*}
    \centering
    \includegraphics[width=0.24\textwidth]{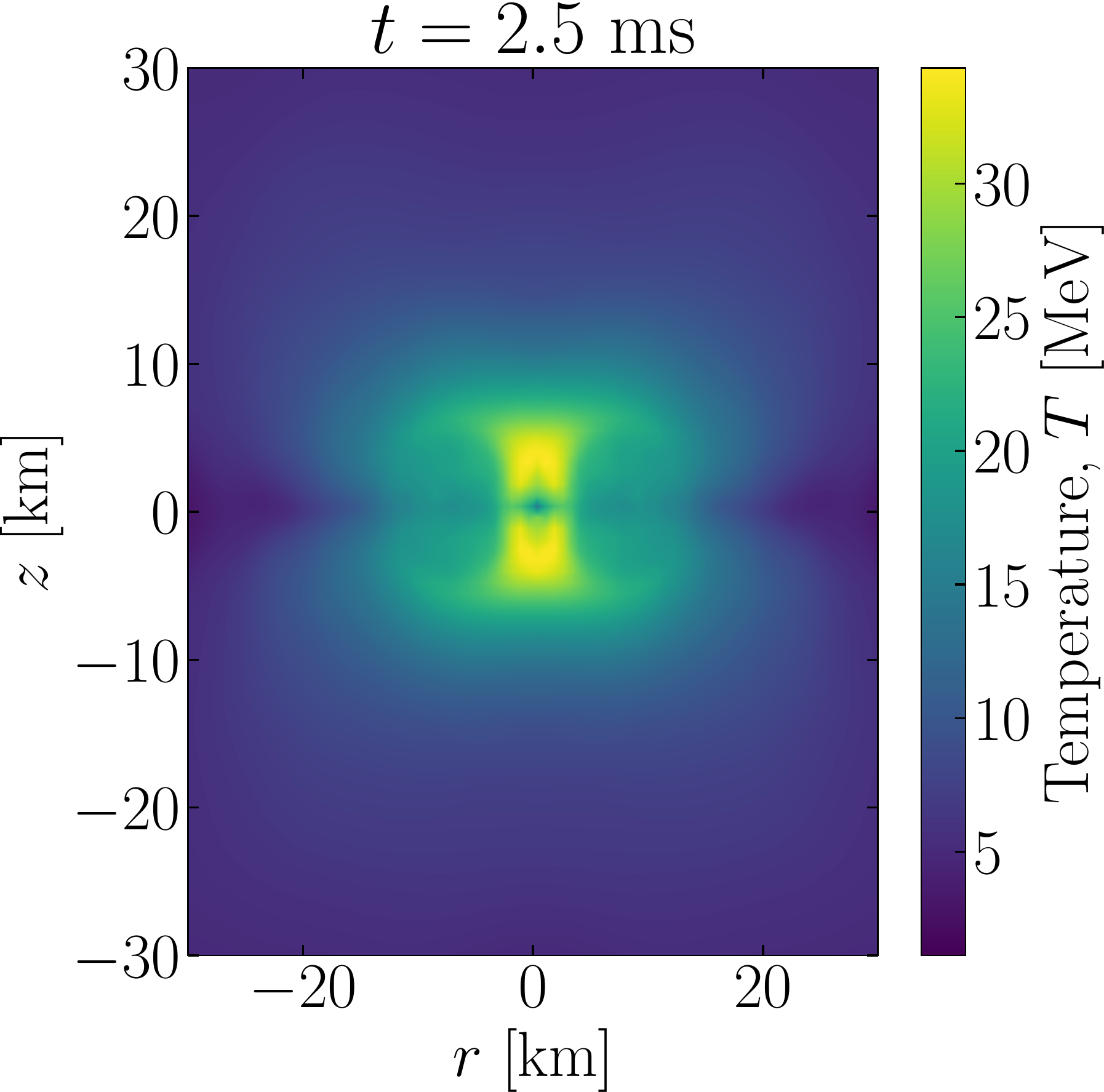}    
    \includegraphics[width=0.24\textwidth]{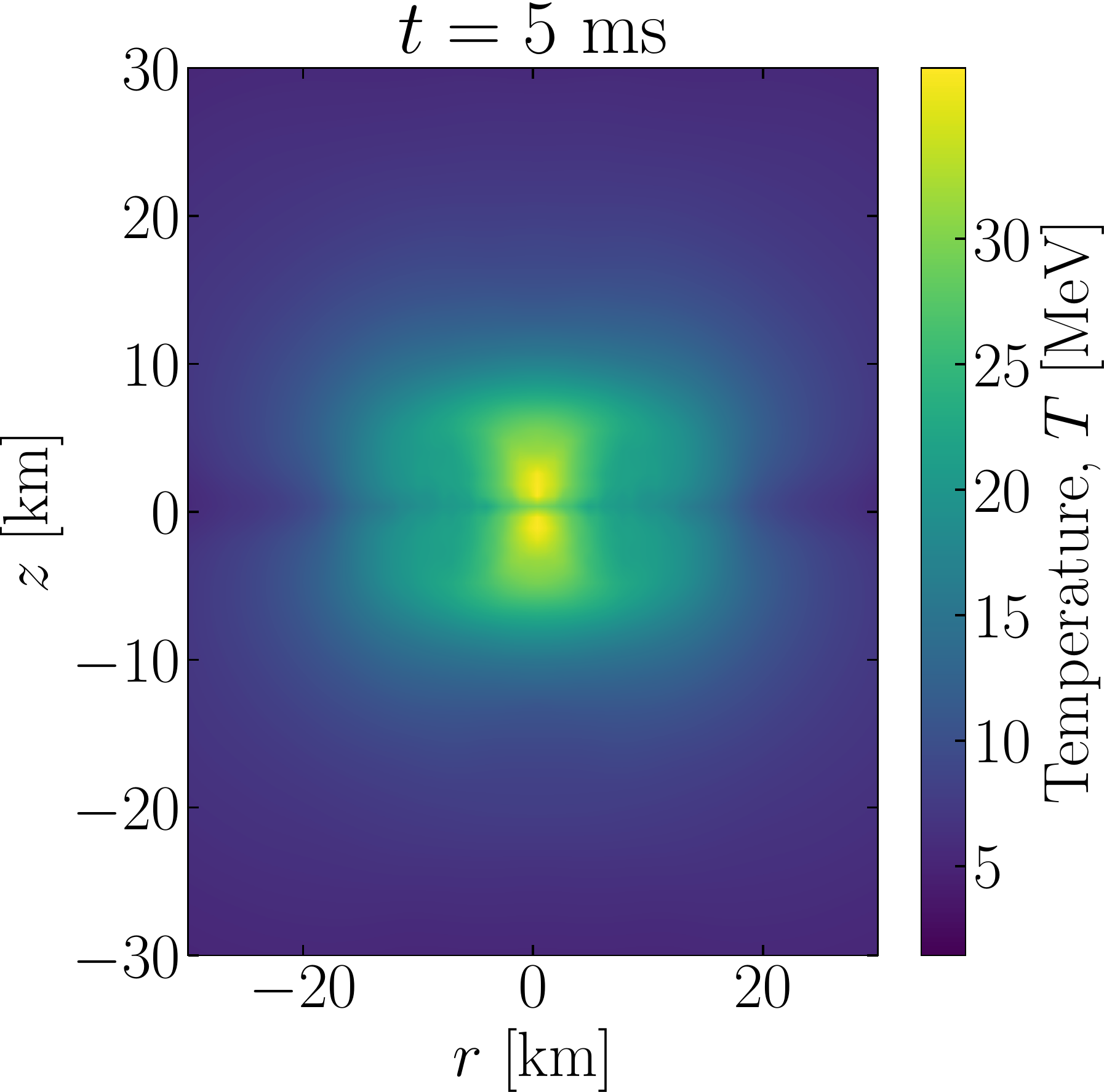}
        \includegraphics[width=0.24\textwidth]{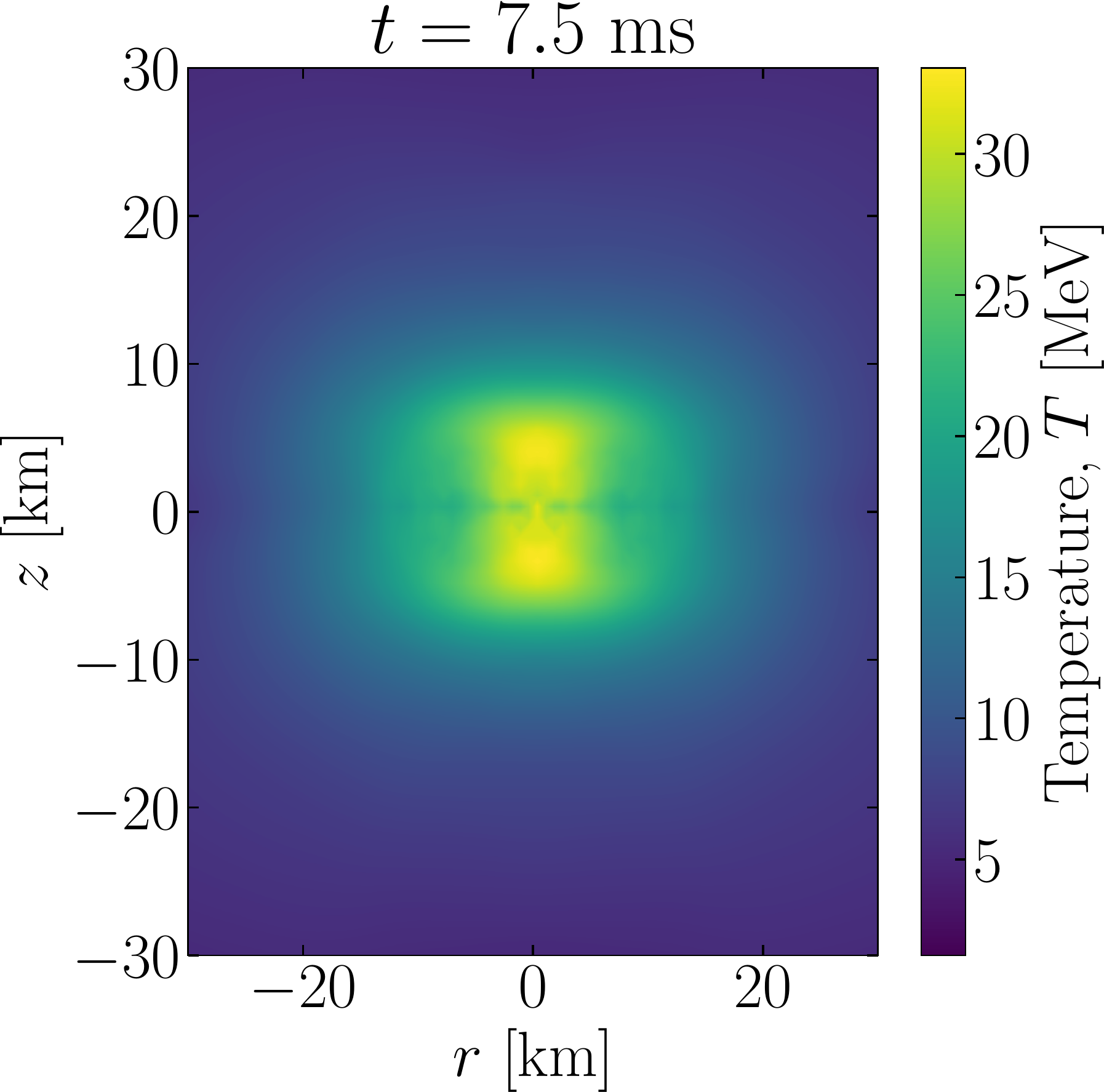}
            \includegraphics[width=0.24\textwidth]{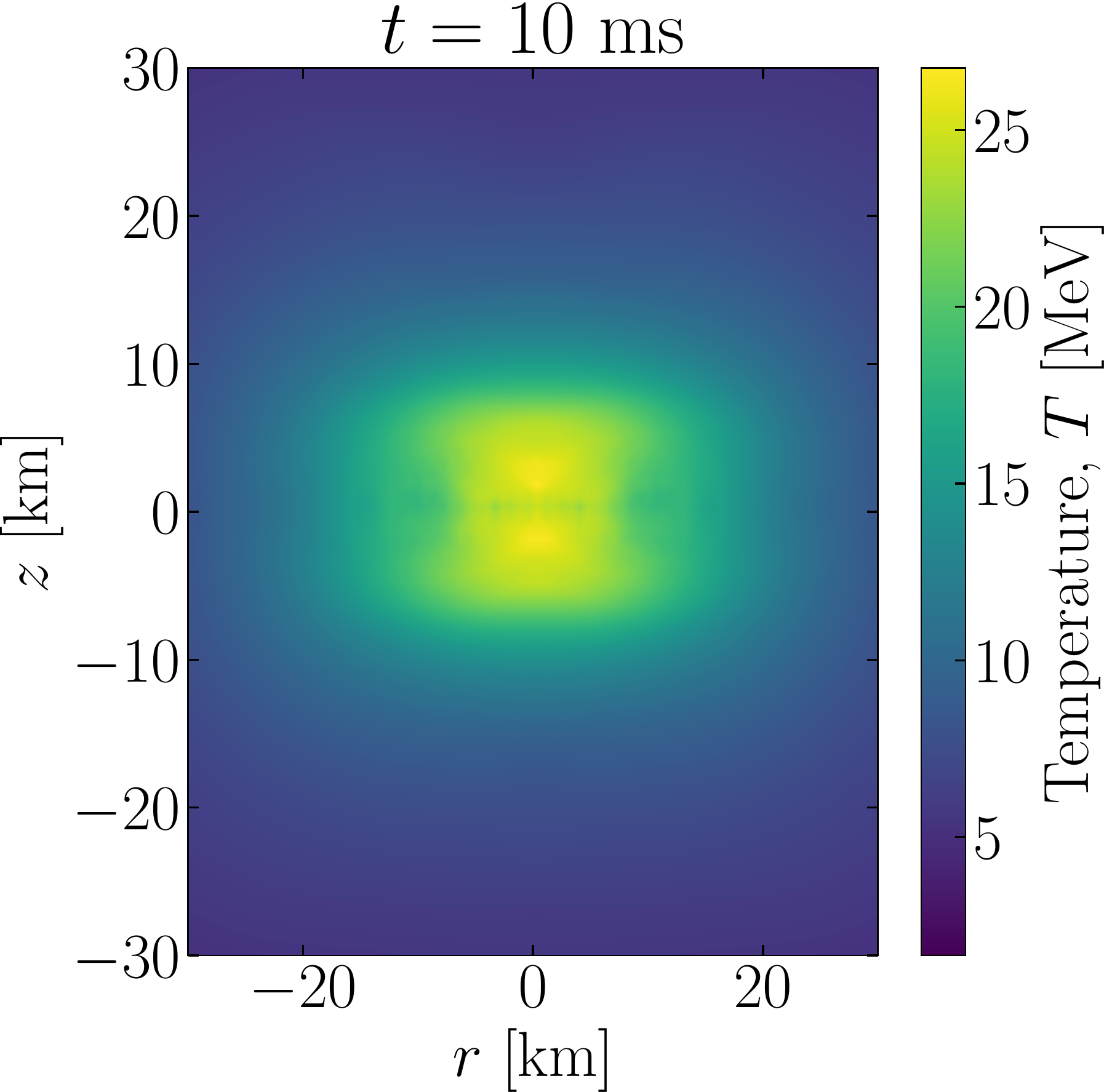}    
            \vskip12pt
  \centering
    \includegraphics[width=0.24\textwidth]{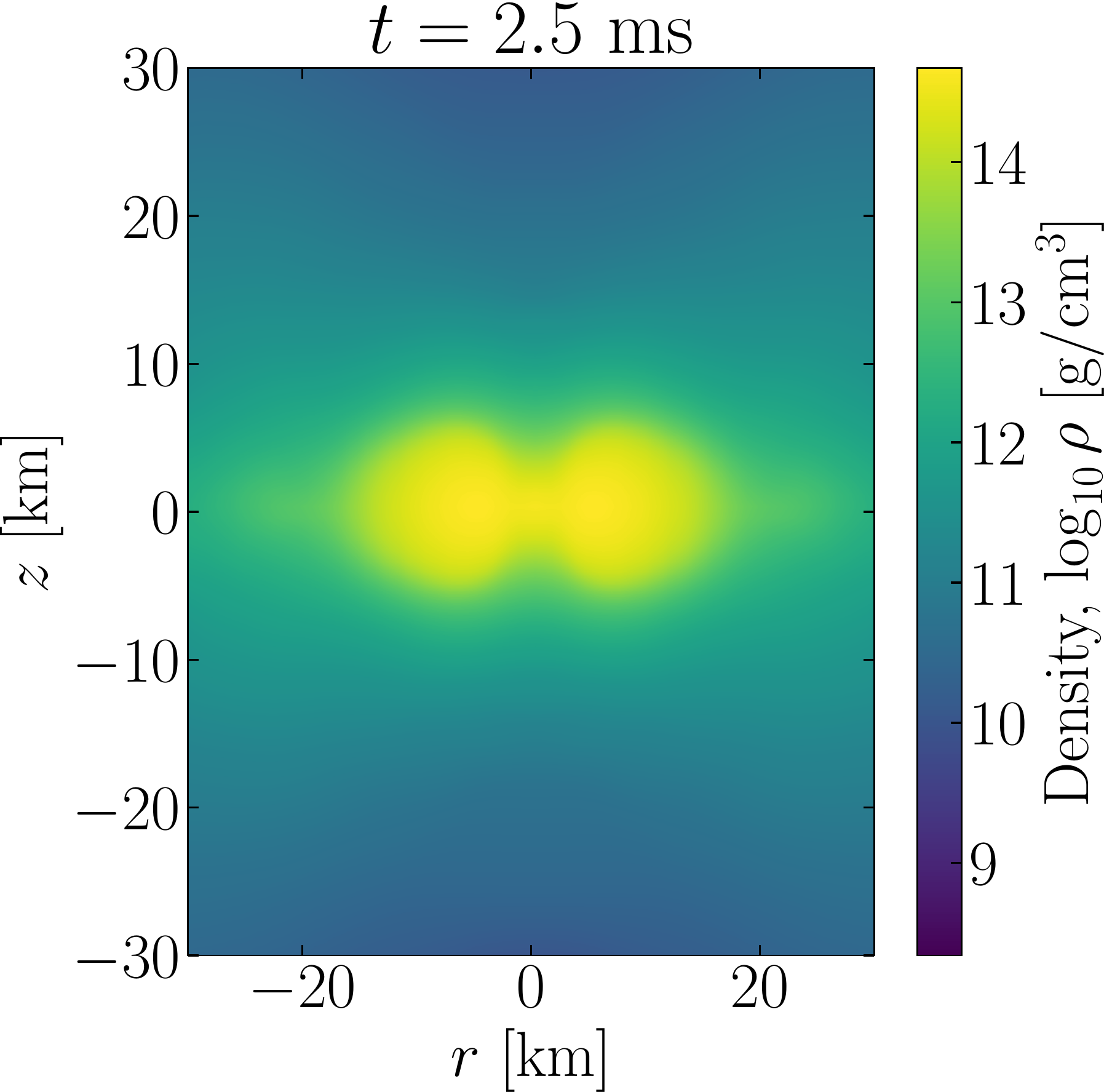}    
    \includegraphics[width=0.24\textwidth]{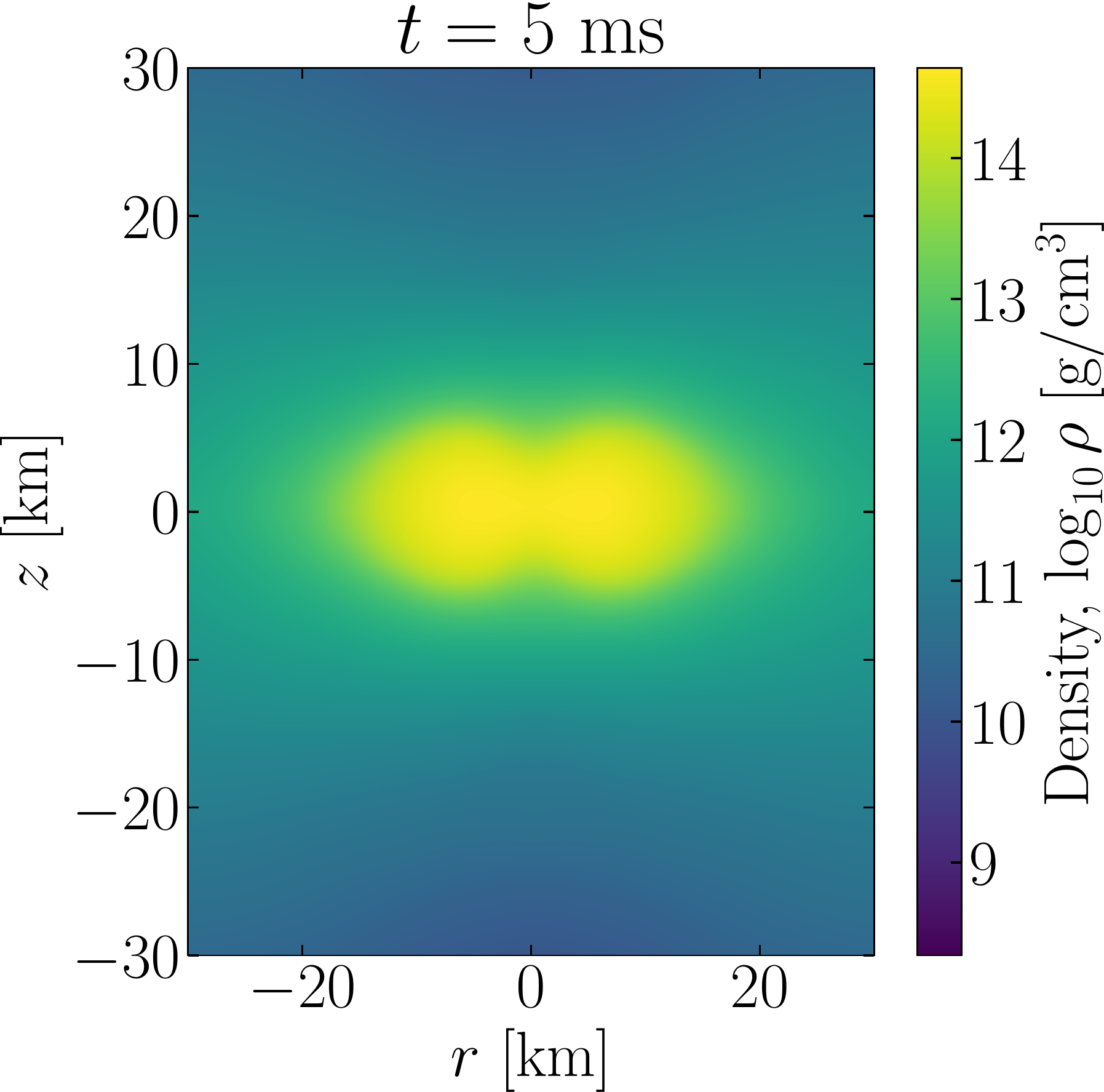}
        \includegraphics[width=0.24\textwidth]{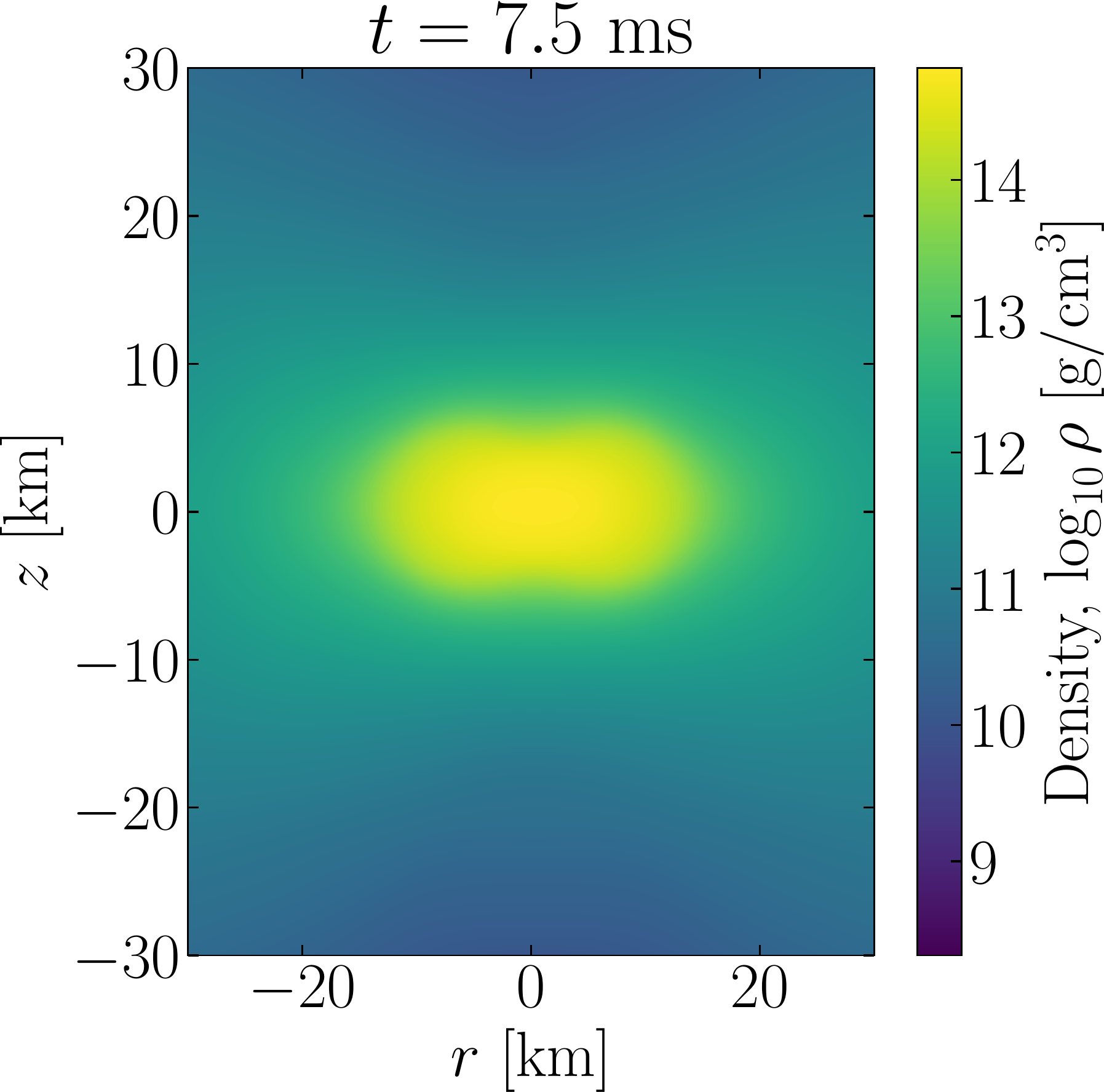}
            \includegraphics[width=0.24\textwidth]{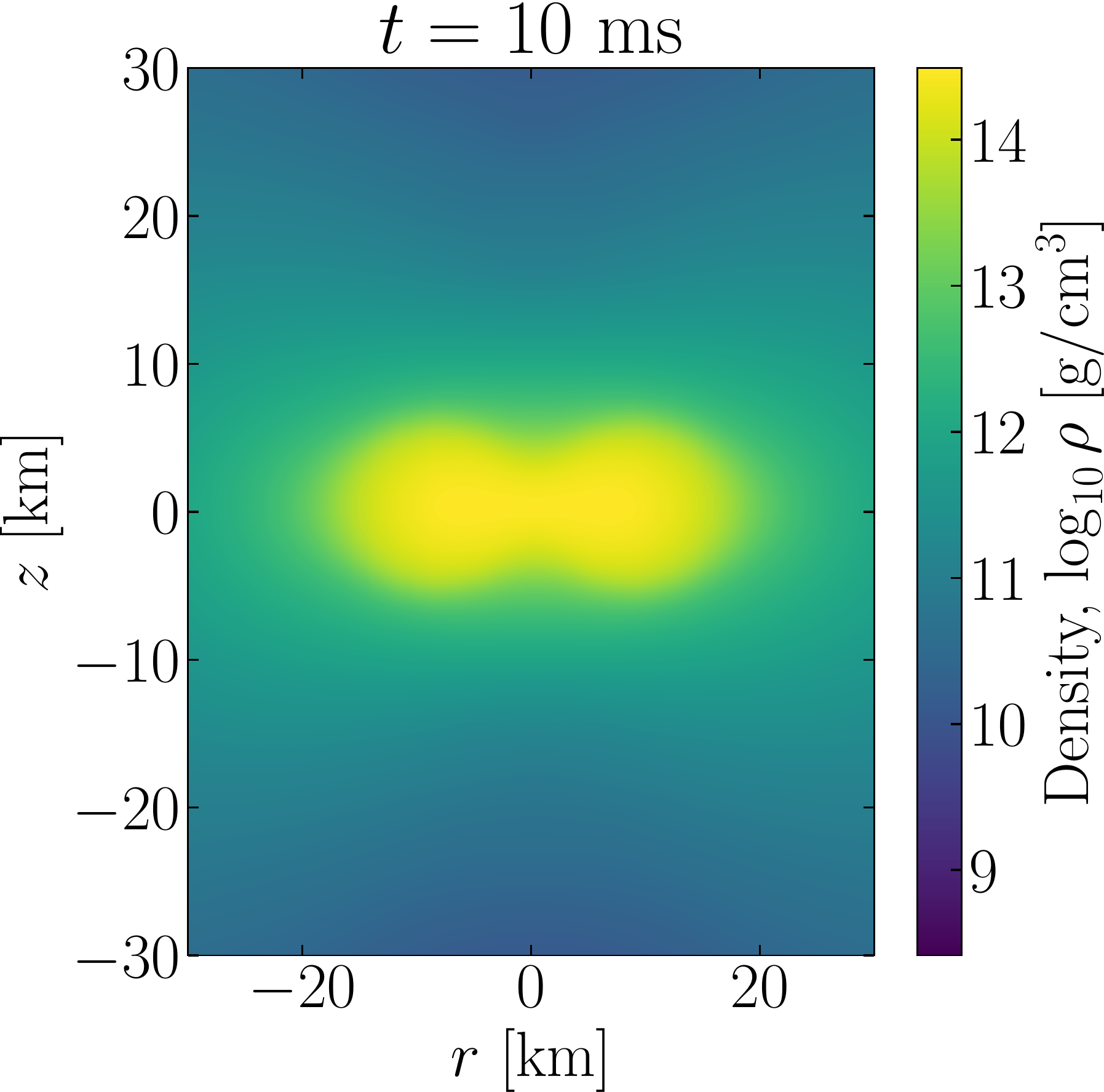}
              \vskip12pt
                  \centering
    \includegraphics[width=0.24\textwidth]{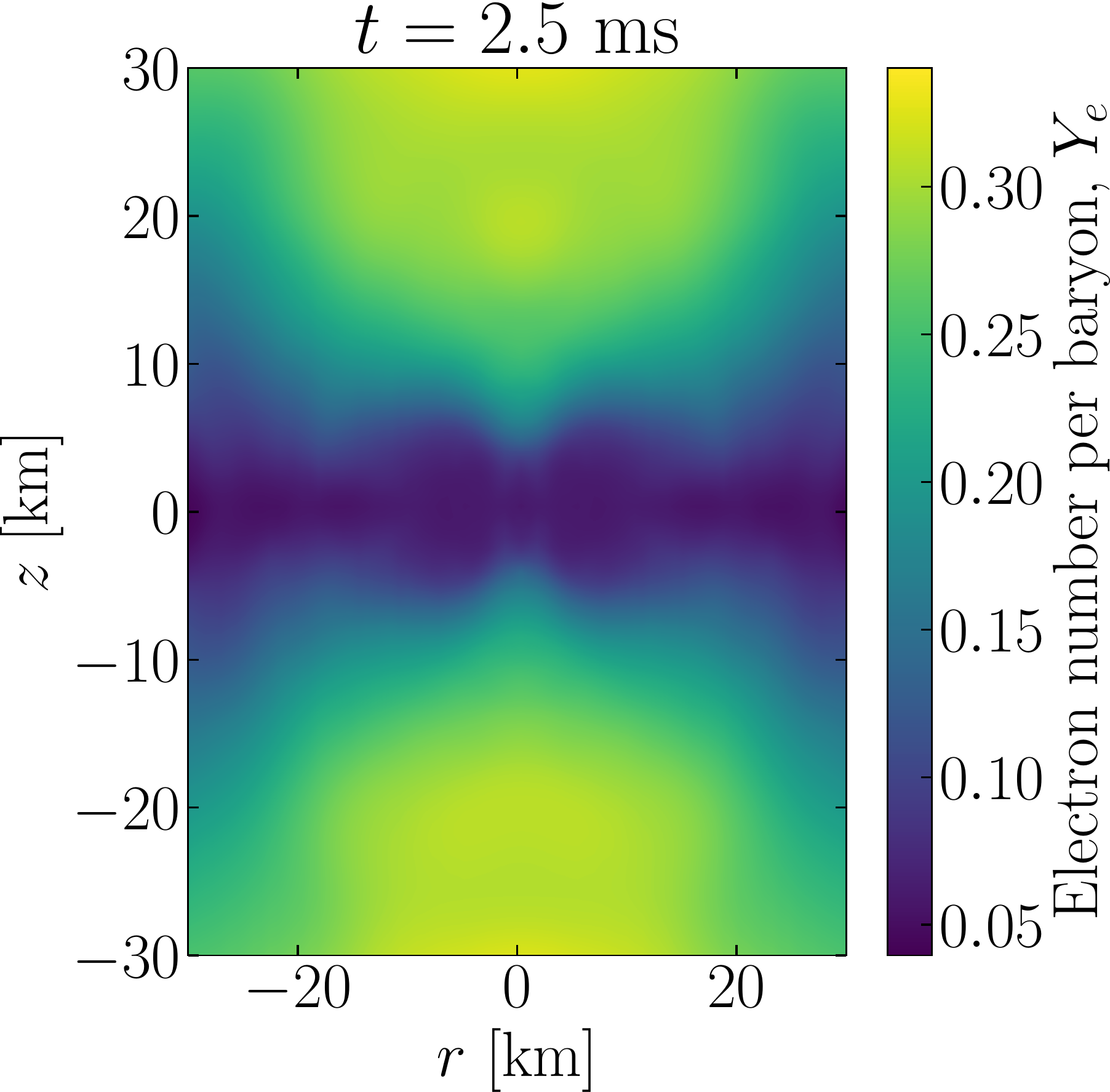}    
    \includegraphics[width=0.24\textwidth]{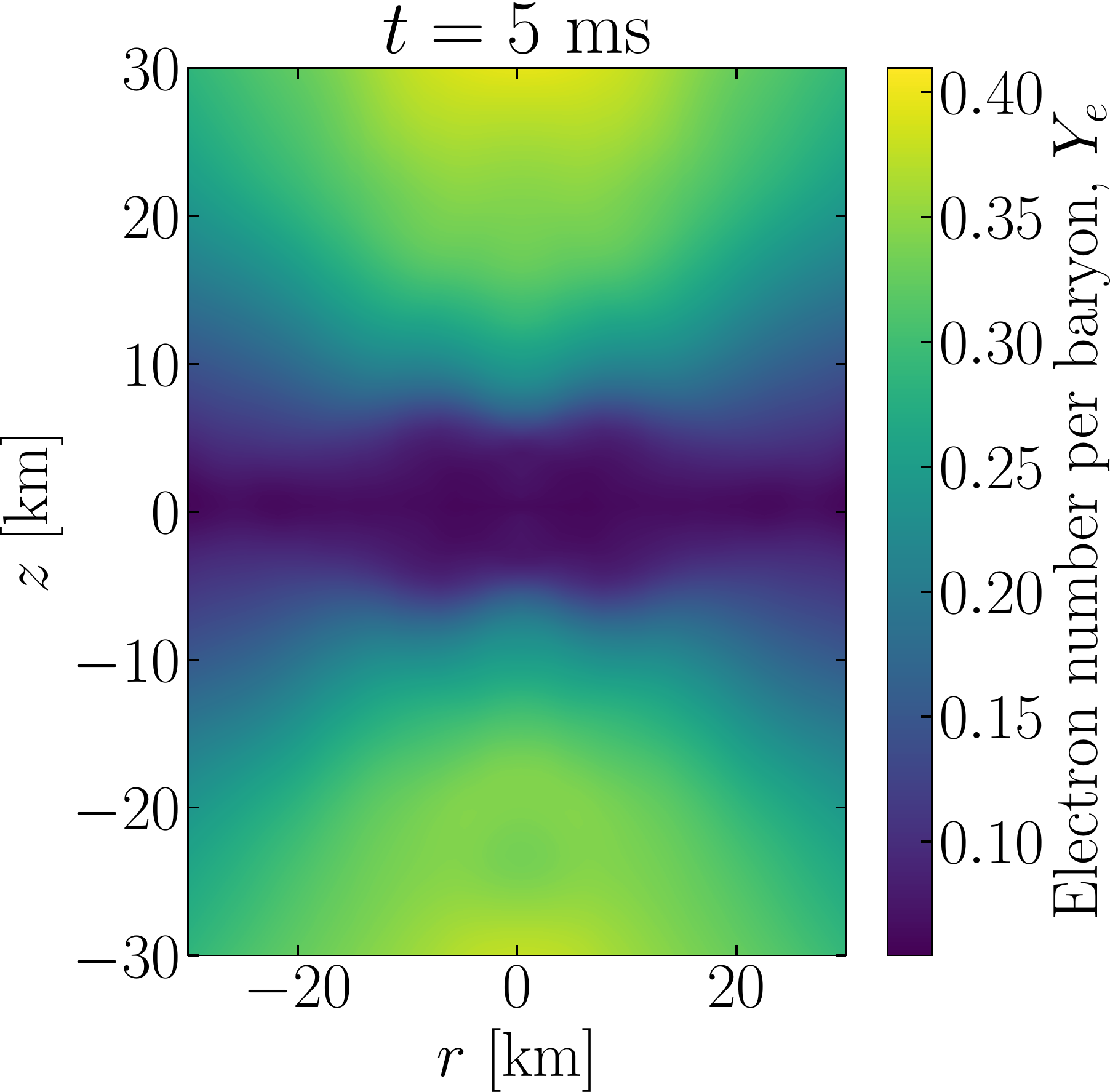}
        \includegraphics[width=0.24\textwidth]{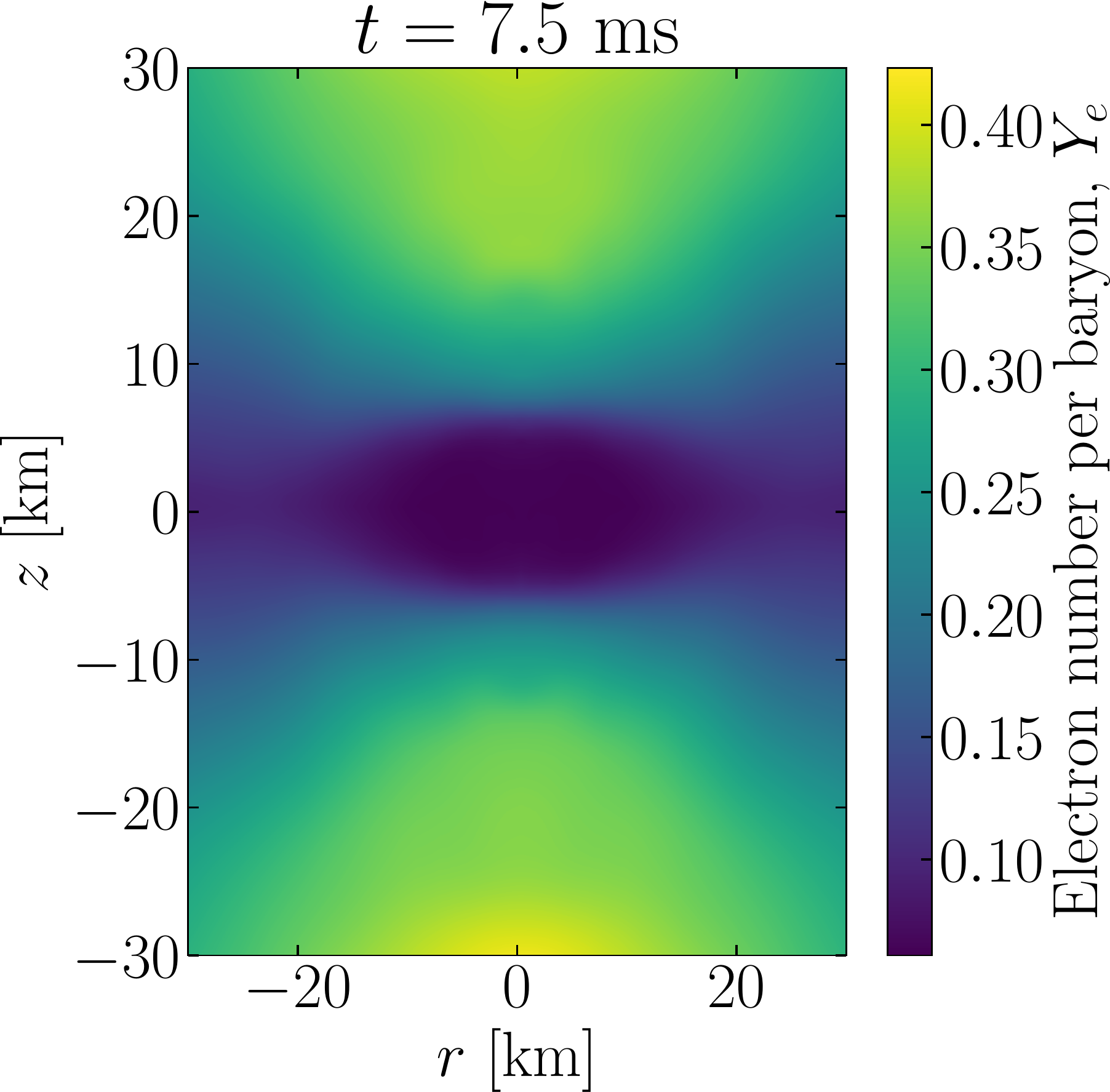}
            \includegraphics[width=0.24\textwidth]{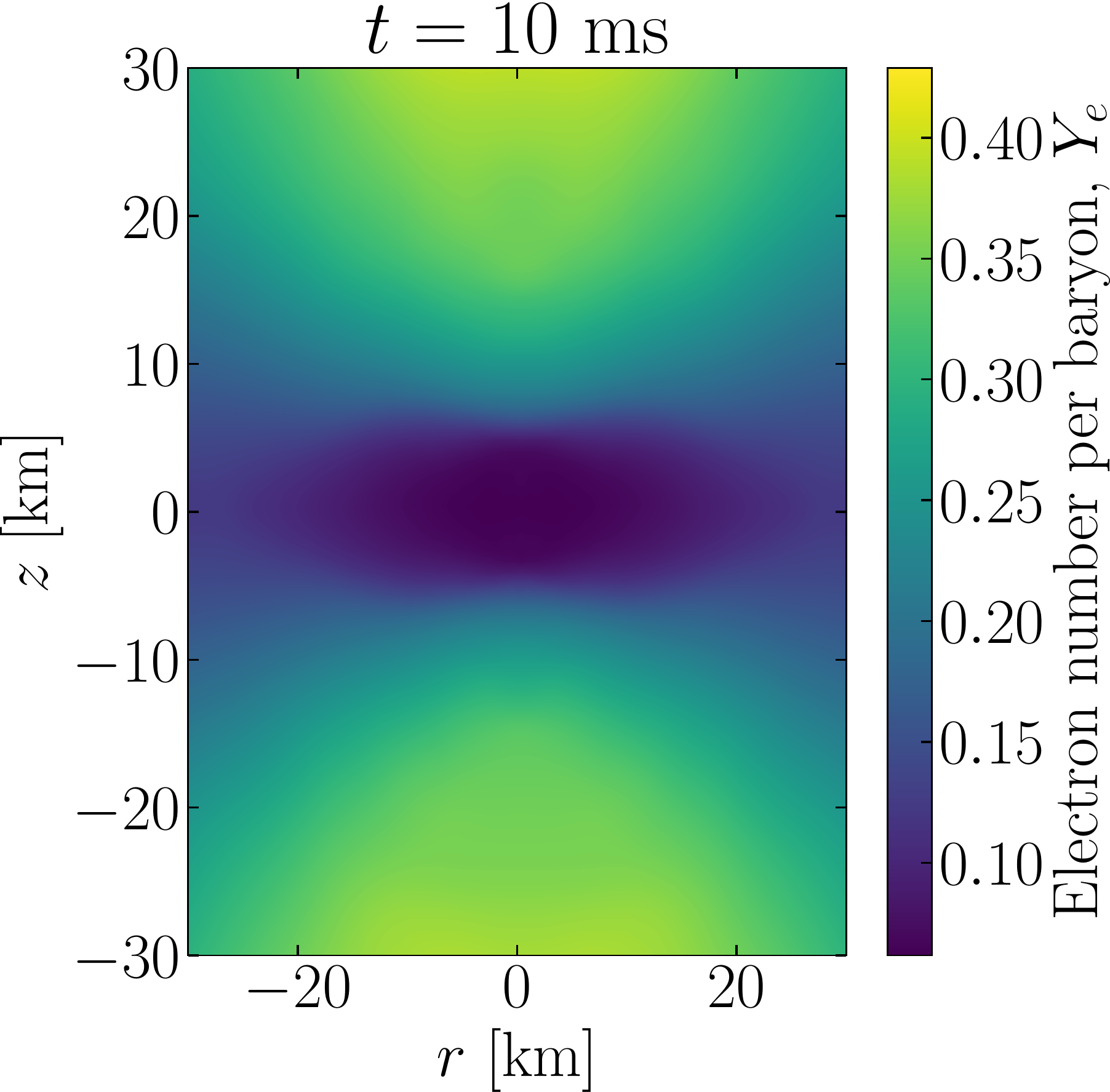}    
            \vskip12pt
  \centering
    \includegraphics[width=0.24\textwidth]{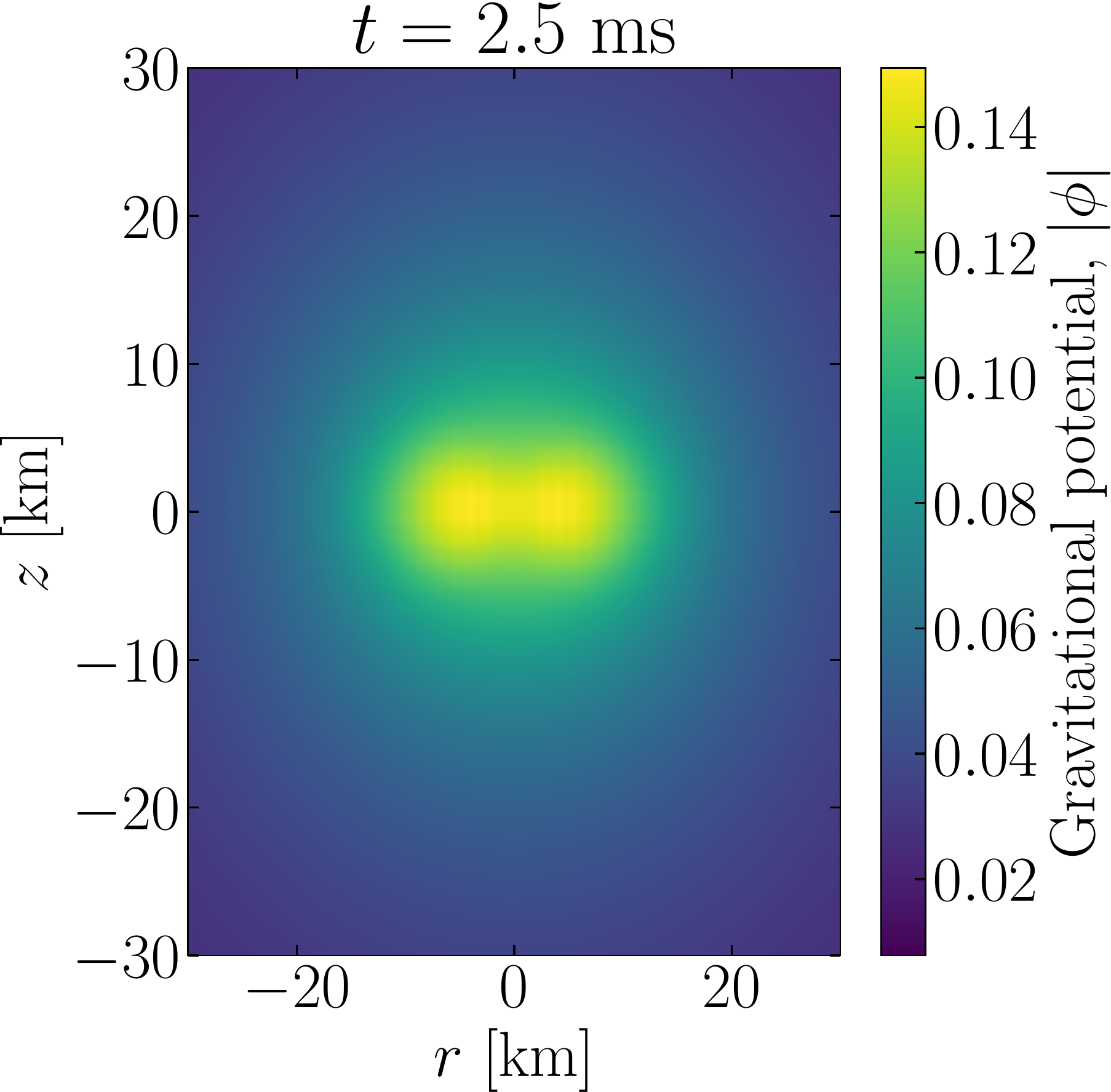}    
    \includegraphics[width=0.24\textwidth]{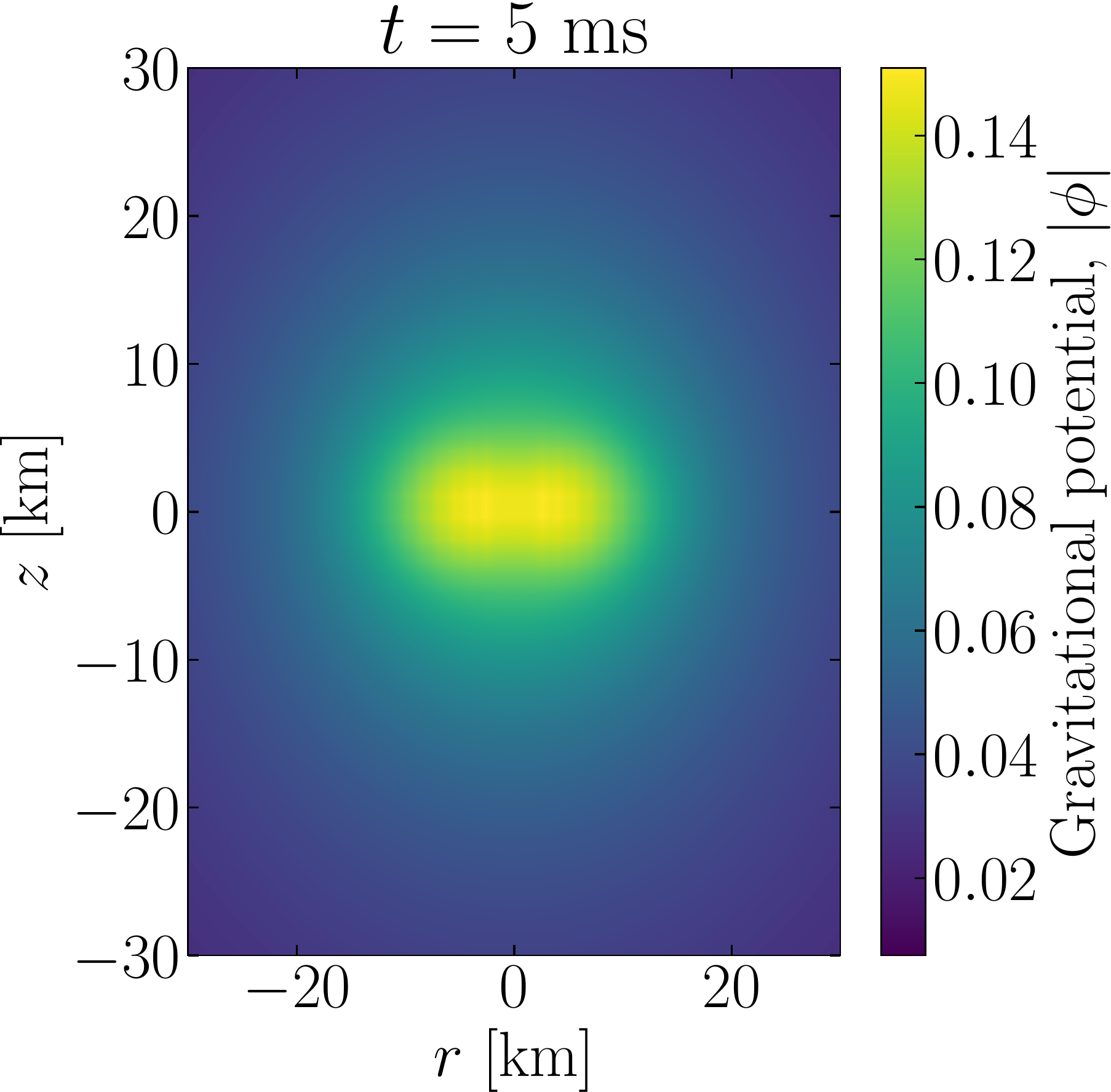}
        \includegraphics[width=0.24\textwidth]{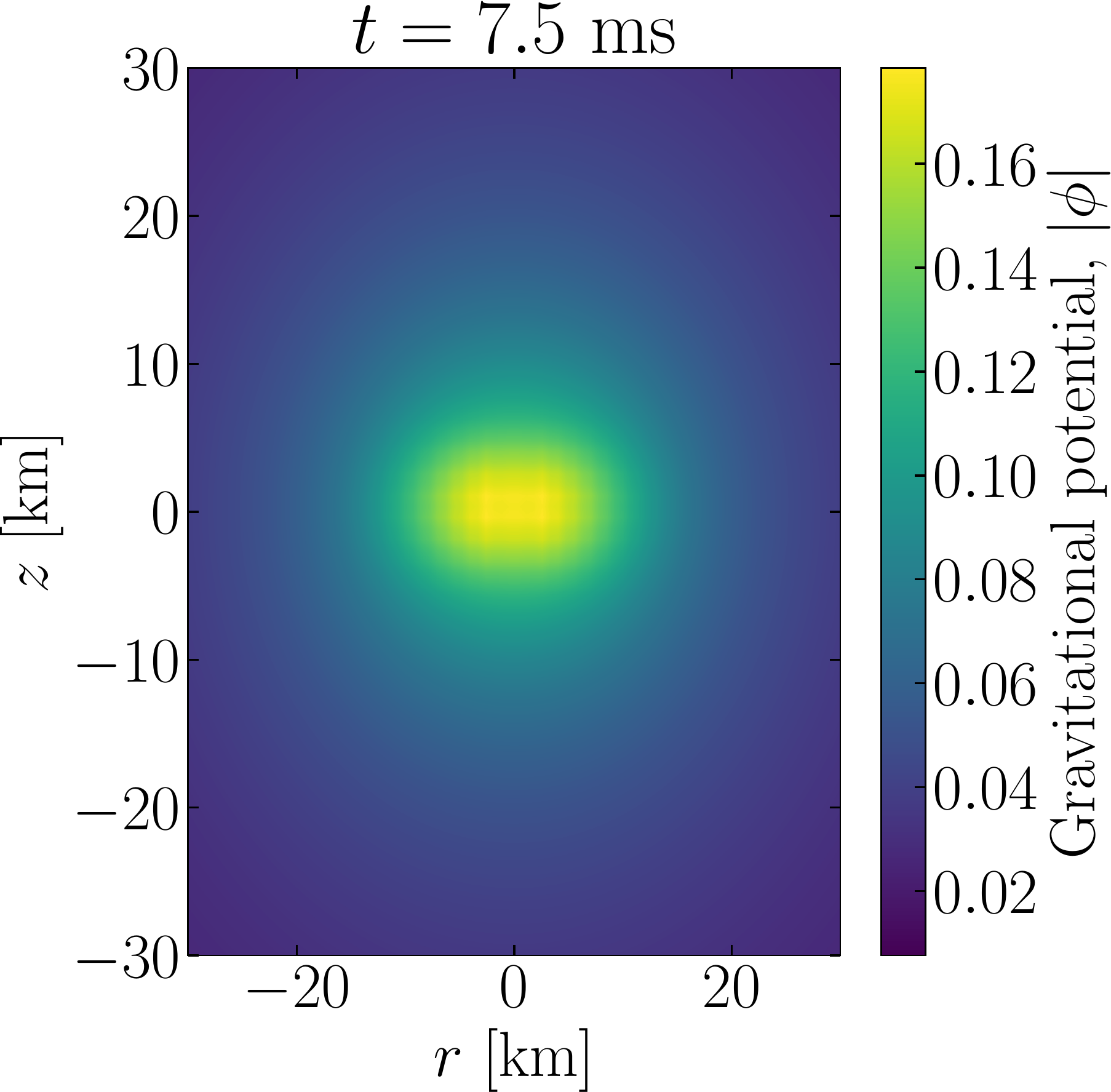}
            \includegraphics[width=0.24\textwidth]{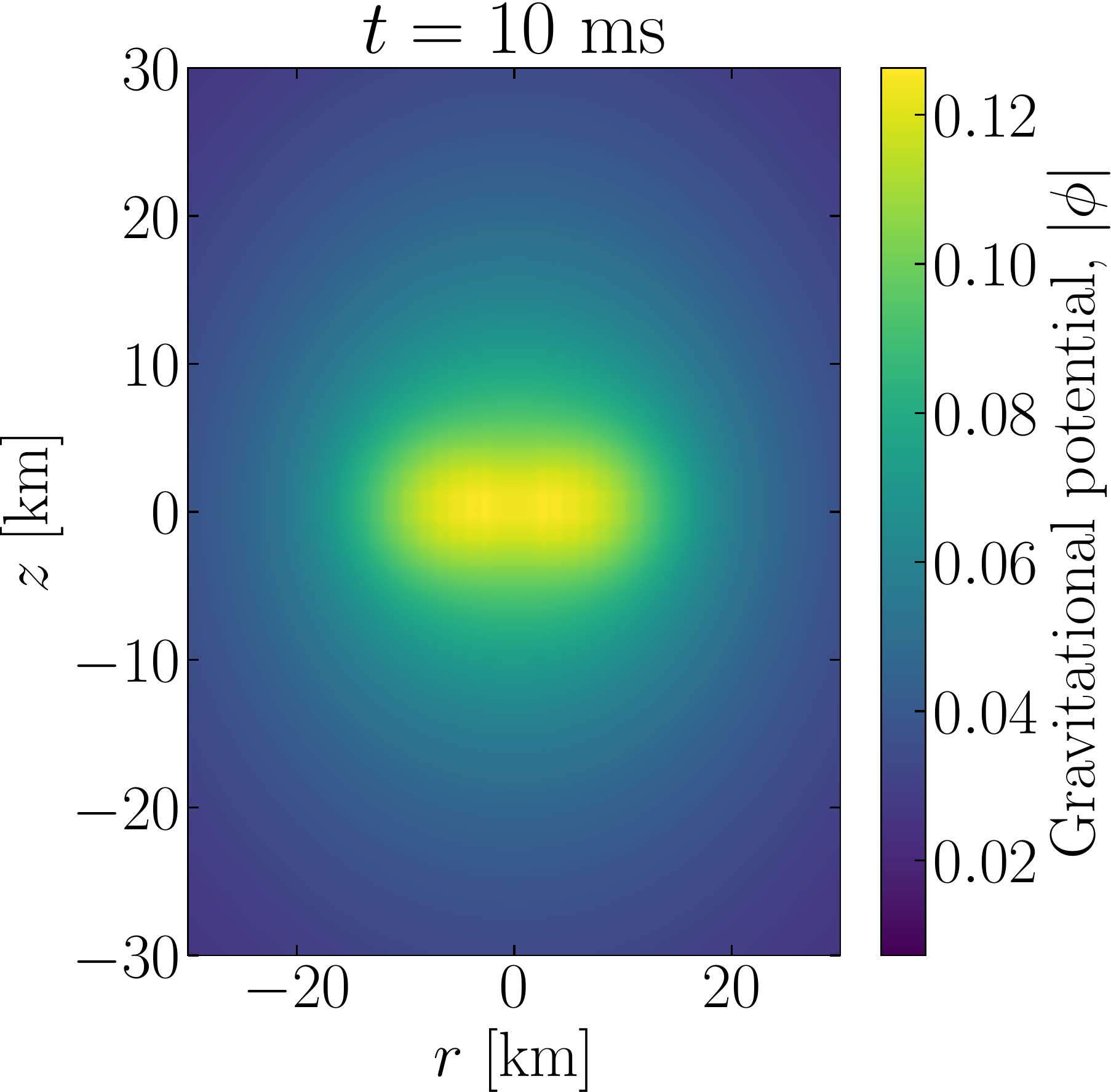}
    \caption{Isocontours of temperature, baryon density, electron number per baryon, and gravitational potential for the NS merger remant simulation with $1.25$ and $1.45\,\rm M_\odot$ NSs and with SFHo EoS~\cite{Ardevol-Pulpillo:2018btx}. 
    }
\vskip12pt
\label{fig:profiles}
\end{figure*}

Equation~(1) of the main text is integrated over a profile model evolving over time. As an example, we show in Fig.~\ref{fig:profiles} isocontours of temperature, baryon density, electron number per baryon, and gravitational potential for our fiducial NS merger remnant model for four snapshots ($2.5$, $5$, $7.5$, and $10\,\rm ms$)~\cite{Bauswein:2013yna,Ardevol-Pulpillo:2018btx,George:2020veu}.

\section{B.~Impact of fast ejecta}

\begin{figure*}
    \centering
    \includegraphics[width=0.95\textwidth]{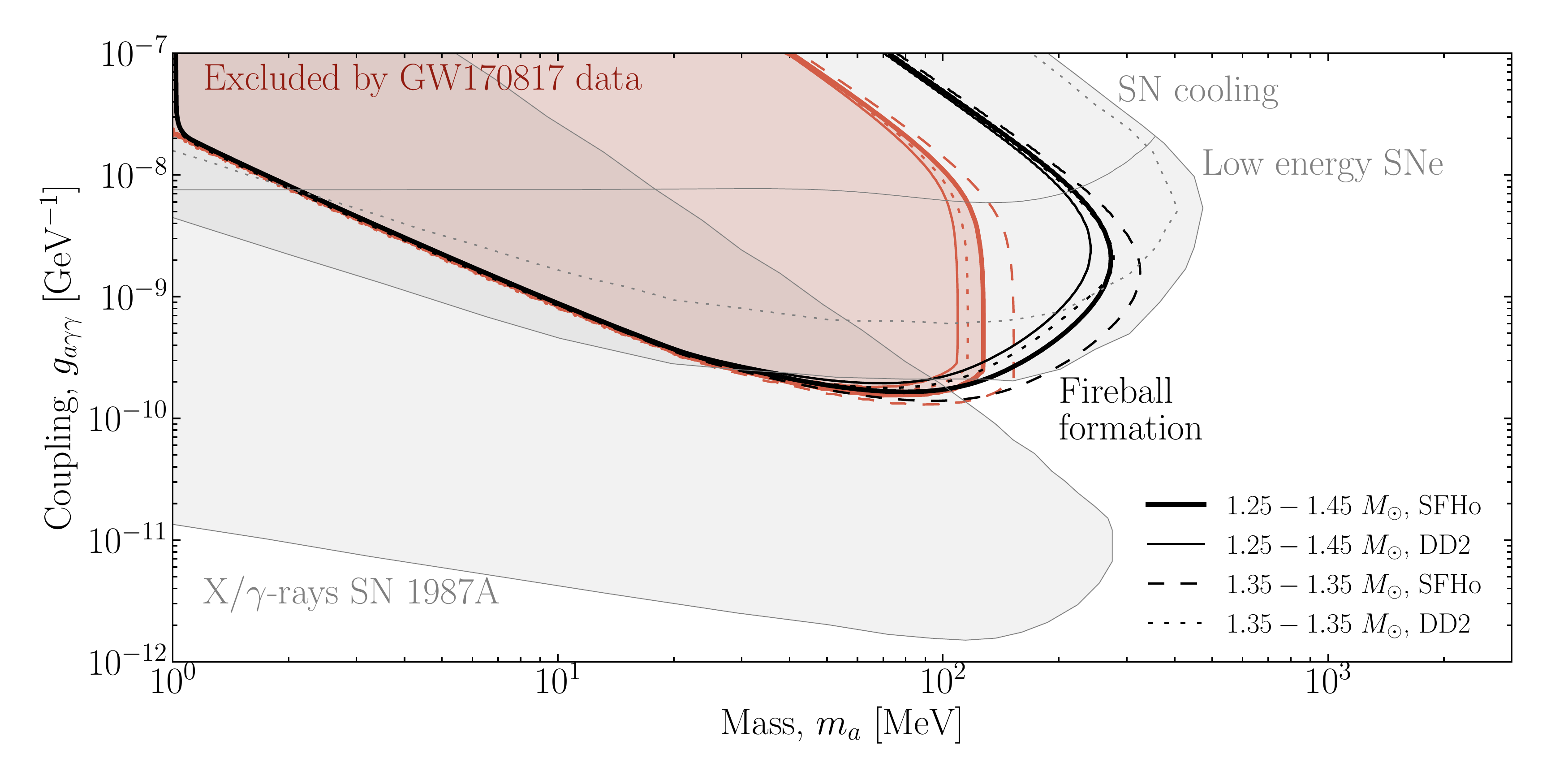}
    \caption{Same as Fig.~3, but  obtained by taking into account for the presence of optically thick ejecta moving with a velocity $V=0.8$. 
    }\label{fig:bounds_ejecta}
\end{figure*}

Fast ejecta with relativistic speeds could be produced in NS mergers, see Ref.~\cite{Shibata:2019wef} for a review.  The escape velocity for the ejecta is of the order of $0.3$ (we use natural units throughout); however, shock heating can accelerate the ejected material to  speeds larger than $0.6$~\cite{Hotokezaka:2012ze}. Axions decaying within the fast ejecta could transfer their energy to the ejecta without reaching Earth. In the main text, we do not consider this possibility and account for energy injection from all the axions decaying outside $1000$~km. In this appendix, we show how our results  change by requiring that only the axions decaying outside the ejecta contribute, which is an overly conservative estimate of the impact of the ejecta.

The main impact of the fast ejecta is to potentially reduce the fraction of energy arising from axion decays going into X-ray and $\gamma$-ray photons. We model the ejecta as a sphere with radius $Vt$, where $V$ is the maximum velocity of the ejecta (see below). The radial position of a photon at the time $t$, assuming collinear emission from the axion decay, can be parameterized as
\begin{equation}
    r_\gamma=v(\omega_a)(t_\mathrm{dec}-t_i)+t-t_\mathrm{dec}\ ,
\end{equation}
where $t_i$ is the time at which the axion is emitted, which we assume to be uniformly distributed between $0$ and $\delta t=1$~s; $t_\mathrm{dec}$ is the time at which the axion decays; $\omega_a$ is the axion energy and $v(\omega_a)$ is the axion velocity. The probability distribution of $t_\mathrm{dec}$ is an exponential,
\begin{equation}\label{eq:prob_dec_fast_ejecta}
    P(t_\mathrm{dec})dt_\mathrm{dec}=e^{-(t_\mathrm{dec}-t_i)/\tau}\frac{dt_\mathrm{dec}}{\tau}\ ,
\end{equation}
where $\tau$ is the energy-dependent lifetime of the axions.

Accounting for the fast ejecta means that we require that only those axions decaying outside of the ejecta, namely those with $v(t_\mathrm{dec}-t_i)>Vt_\mathrm{dec}$ must be included. In addition, we must exclude the axions decaying inside a radius of about $R=1000$~km. The probability that an axion satisfies both conditions is
\begin{equation}
P_\mathrm{dec}=\frac{1-e^{-\mu}}{\mu}e^{-R/v\tau}\theta(v-V)\ ,
\end{equation}
where
\begin{equation}
\mu=\frac{V \delta t}{(v-V)\tau}
\end{equation}
and $\theta(x)$ is the Heaviside function. Therefore, the total energy $\mathcal{E}$, particle number $\mathcal{N}$, and radial momentum $\mathcal{P}$ injected by axion decay must include this additional factor which suppresses the contribution of slow axions decaying at small radii. This is especially relevant at large masses, where the production at large energies is exponentially suppressed by the Boltzmann factor; because of the fast ejecta, only those particles with an energy larger than $m_a/\sqrt{1-V^2}$ are actually contributing to the flux.

In addition to reducing the total energy injected in the fireball, the presence of fast ejecta also slightly changes the average radius and thickness of the photon shell. As in Ref.~\cite{Diamond:2023scc}, we obtain these quantities as the average of $r=\langle r_\gamma(\bar{t})\rangle$ and $\Delta=\sqrt{\langle r_\gamma(\bar{t})^2\rangle-\langle r_\gamma(\bar{t})\rangle^2}$, where the average is performed over the energy distribution and the probability distribution of the decay time $t_\mathrm{dec}$ and the time of emission $t_i$, and $\bar{t}=\langle t_\mathrm{dec}\rangle$ is the average time of axion decay. We do account for these changes, although they have very limited impact compared to the reduction in the total energy injected by axion decay.

The key unknown quantity here is the velocity of the ejecta. Most of the material expelled from the merger has a typical velocity $0.3$ of the order of the escape velocity at the radius of dynamical ejection. However, a small percentage of mass, which various simulations report between $10^{-6}$~and $10^{-4}$~$M_\odot$~\cite{Radice:2018pdn,Kullmann:2021gvo} is expelled with faster velocities, in excess of $0.6$. This material would be more than enough to trap the photons by Thomson scattering; in fact, using the Thomson cross section, one easily finds that photons produced at a radius of about $1$~s would be trapped by an amount of material as small as $10^{-11}$~$M_\odot$. Clearly, the critical parameter for our purposes is the maximum speed with which the ejecta are released. This is however a poorly known parameter from numerical simulations, see, e.g., Ref.~\cite{Radice:2018pdn}. References~\cite{Hotokezaka:2012ze,Radice:2018ghv} claim a maximum asymptotic speed for the ejecta of about~$0.8$.
Therefore, for our purposes, we use $V=0.8$~as our benchmark value.

Figure~\ref{fig:bounds_ejecta} shows the bounds obtained with this reference choice. The bounds are obviously weaker than the ones shown in the main text, since only the axions moving with a speed larger than $V$  contribute to  fireball formation. The excluded region has a clear vertical cut; this can be easily understood, since the total energy injected into axions increases as $g_{a\gamma\gamma}^2$, but only a fraction of the axions of the order of $\tau/\Delta\propto g_{a\gamma\gamma}^{-2}$ are produced early enough that they can overcome the fast ejecta (see Eq.~\eqref{eq:prob_dec_fast_ejecta}), so the total energy visible at Earth is independent of the coupling. The results show that even accounting for the fast ejecta the bounds are still competitive with the bounds set from low-energy SNe, and are stronger than the most conservative version of the low-energy SNe bound (gray dotted line) obtained assuming a total energy in axions from SNe smaller than $10^{51}\,\rm erg$. One should also keep in mind the critical interplay between fast ejecta and the  temperature of the HMNS core; for a larger temperature, axions with relativistic speeds are more easily produced and can therefore escape the ejecta in larger quantities. As the one-zone model illustrates, the axion emission is suppressed by a factor $e^{-m_a/T\sqrt{1-V^2}}$, where the factor $\sqrt{1-V^2}$ appears since only axions faster than $V$ can decay outside of the ejecta. This shows that a temperature even a factor $2$ higher could compensate the enhanced exponential suppression from the factor $\sqrt{1-V^2}\simeq 0.6$ for $V=0.8$. The profiles used in Ref.~\cite{Camelio:2020mdi} have peak temperatures even a factor $3$ or higher than the ones used for the results shown in the main text, therefore for such hotter profiles the bounds would likely close at higher masses and still exclude sizable new regions of parameter space.

\section{C.~Impact of uncertainties in the merger simulation}

The simulations we have used to obtain our benchmark bounds are based on two choices of masses of the neutron stars ($1.35$--$1.35$~$M_\odot$ and $1.25$--$1.45$~$M_\odot$) and two choices of EoS (SFHo and DD2). In the main text, we have shown through a one-zone model how  only a few properties of the HMNS formed in the merger affect the bounds, namely the typical temperature and size of the HMNS, and its lifetime. Here we discuss in  detail how these quantities are related to the masses of the NSs and the choice of EoS, clarifying the assumptions on which our bounds rely.

A source of uncertainty comes from the assumed EoS. In this work, we use   stiff and soft EoSs,  DD2 and SFHo, that should braket any other choice of EoS. As a general trend, a stiff EoS leads to relatively lower temperatures, due to the collapse happening at relatively larger radii. We may expect any bound linked to a different EoS to lie in-between the benchmark cases we show in the main text. We should  emphasize that the uncertainty connected to the EoS similarly affects  the complementary bounds from supernova observations; therefore, a complementary astrophysical objects reinforcing exclusions in this region with independent measurements is all the more relevant. 

The mass ratio of the colliding NSs also has an impact on the strength of the  obtained bounds, as our figures in the main text illustrate. The region of fireball formation and the bounds at low masses, below  $m_a\simeq 100$~MeV, are negligibly affected. Our bounds are competitive with the supernova bounds in this region, without a strong dependence on the choice of the NS masses. Above $m_a\gtrsim 100$~MeV, the bounds and region of fireball formation depend more sensitively on the masses. For our chosen models, the asymmetric merger  shows relatively lower temperatures, and thus lead to slightly weaker bounds. However, we stress that the effect depends on the EoS. For example, the DD2 EoS shows extremely little dependence on the mass ratio, at the level of tens percent and smaller. The reason is that, for the most conservative case of a very stiff EoS, the typical temperature scale reached in the remnant  never falls far below about $15$~MeV. 

In principle, the range of NS masses implied by the GW measurements from LIGO is from $1.0\;M_\odot$ to $1.9\; M_\odot$ assuming a high-spin prior---no evidence for high spin has however been found by LIGO---or from $1.16\; M_\odot$ to $1.9\; M_\odot$ for low-spin prior. The total mass of the two NSs is estimated to be $2.7\;M_\odot$. Testing the entire mass range implied by these observations is beyond the scope of this work. However, we  remark that a large degree of asymmetry in the masses of the two NSs may lead to prompt collapse~\cite{Nedora:2020hxc}. As we discuss in the main text,  the kilonova and the gamma-ray burst coincident with GW170817  point to the conclusion  that the remnant  survived for a certain amount of time, likely between hundreds of ms and $1$~s (see, e.g., Refs.~\cite{Gill:2019bvq,Bauswein:2017vtn}). The survival of the merger 
remnant as a HMNS for some time (in contrast to prompt collapse)
is generally accepted, and prevents us from using currently available simulations of high-mass ratio mergers.

\begin{figure*}
    \includegraphics[width=\textwidth]{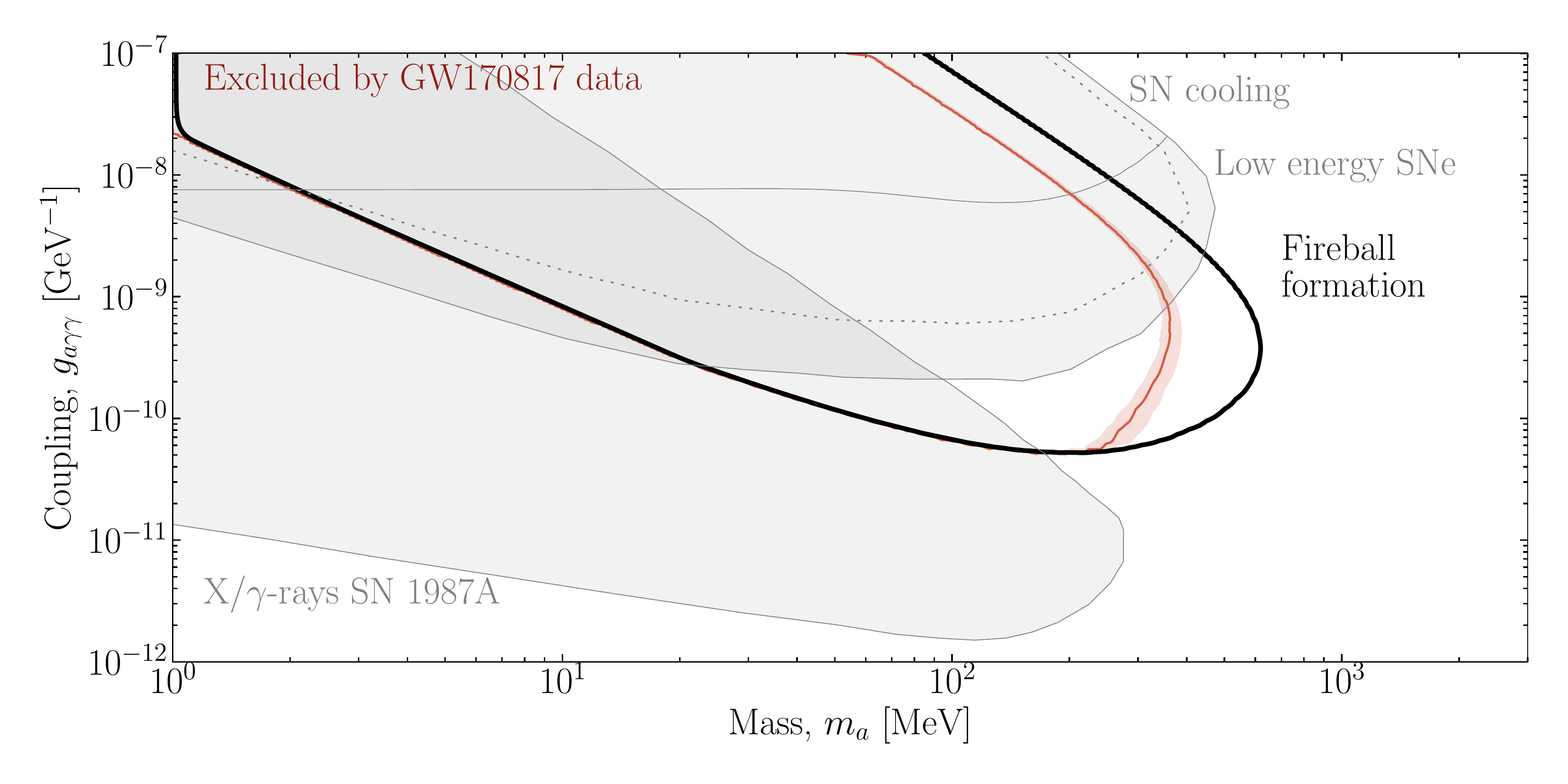}
    \caption{Impact of distance uncertainty on the bounds, for the model $1.25 - 1.45\;M_\odot$, SFHo.}\label{fig:bounds_distance}
\end{figure*}

Finally, the distance of the merger from the Earth is itself affected by uncertainties. The LIGO collaboration estimates this distance to be $d_L=40^{+8}_{-14}$~Mpc~\cite{LIGOScientific:2017vwq}. The uncertainty in the distance of course does not affect the region of fireball formation, which depends only on the local properties of the merger. On the other hand, it can affect the lateral closure of our bounds, determined by the energy fluence observed at Earth. Figure~\ref{fig:bounds_distance} shows the impact on the bounds by taking the full uncertainty interval for this distance; since the energy fluence simply scales as $d_L^{-2}$, the scaling dependence on the distance can also be estimated directly from our expressions for the one-zone model, giving $g_{a\gamma\gamma}\propto d_L$, implying a $20-35\%$ uncertainty in the bounds coming from the distance uncertainty. Notice that the electromagnetic counterpart of GW170817 allows a distance measurement to a better precision of about $10\%$~\cite{Hjorth:2017yza}. In any case, as we have shown, even the localization purely by gravitational wave observations does not significantly affect the bounds.

\end{document}